\documentclass[aps,prl,reprint,groupedaddress]{revtex4-1}

\usepackage{amsmath}
\usepackage{mathtools}
\usepackage{graphicx}
\usepackage{upgreek}
\usepackage{graphicx}
\usepackage[titletoc]{appendix}
\usepackage{xcolor}
\definecolor{prlblue}{rgb}{0.176, 0.152, 0.57}
\usepackage[colorlinks=true, linkcolor=black, citecolor=prlblue, filecolor=black, urlcolor=prlblue]{hyperref}
\usepackage[draft]{changes}
\usepackage[english,capitalise]{cleveref}
\definechangesauthor[name={Thomas}, color=orange]{T}
\usepackage{lineno}
\usepackage[compact]{titlesec}
\titlespacing{\section}{0pt}{*0}{*0}
\titlespacing{\subsection}{0pt}{*0}{*0}
\titlespacing{\subsubsection}{0pt}{*0}{*0}

\begin{document}

\title{All-optical density downramp injection \\in  electron-driven plasma wakefield accelerators}

\author{D. Ullmann$^{1,2}$, P. Scherkl$^{1,2}$, A. Knetsch$^3$, T. Heinemann$^{1,2,3,4}$, A. Sutherland$^{1,2,5}$, A. F. Habib$^{1,2}$, O. S. Karger$^{4}$, A. Beaton$^{1,2}$, G. G. Manahan$^{1,2}$, A. Deng$^{6}$, G. Andonian$^{6, 8}$, M. D. Litos$^{7}$, B. D. O'Shea$^{5}$, D. L. Bruhwiler$^{9}$, J. R. Cary$^{7,11}$, M. J. Hogan$^{5}$, V. Yakimenko$^{5}$, J. B. Rosenzweig$^{6}$, and B. Hidding$^{1,2}$}

\affiliation{$^1$SUPA, Department of Physics, University of Strathclyde, UK. $^2$The Cockcroft Institute, Daresbury, UK. $^3$Deutsches Elektronen-Synchrotron DESY, Hamburg, Germany. $^4$	Department of Experimental Physics, University of Hamburg, Hamburg, Germany. $^5$SLAC National Accelerator Laboratory, Menlo Park, California, USA. $^6$Department of Physics and Astronomy, University of California Los Angeles, USA. $^7$Center for Integrated Plasma Studies, Department of Physics, University of Colorado, Boulder, Colorado, USA. $^8$Radiabeam Technologies, Santa Monica, CA 90404, USA. $^9$RadiaSoft LLC, Boulder, CO 80301, USA.  $^{10}$Tech-X UK Ltd., Daresbury, UK.  $^{11}$Tech-X Corporation, Boulder, USA.}

\date{\today}

\begin{abstract}

Injection of well-defined, high-quality electron populations into plasma waves is a key challenge of plasma wakefield accelerators. Here, we report on the first experimental demonstration of plasma density downramp injection in an electron-driven plasma wakefield accelerator, which can be controlled and tuned in all-optical fashion by mJ-level laser pulses. The laser pulse is directed across the path of the plasma wave before its arrival, where it generates a local plasma density spike in addition to the background plasma by tunnelling ionization of a high ionization threshold gas component. This density spike distorts the plasma wave during the density downramp, causing plasma electrons to be injected into the plasma wave. By tuning the laser pulse energy and shape, highly flexible plasma density spike profiles can be designed, enabling dark current free, versatile production of high-quality electron beams. This in turn permits creation of unique injected beam configurations such as counter-oscillating twin beamlets.

\end{abstract}


\maketitle

\section{I. INTRODUCTION}
In electron beam-driven \cite{Chen1985PWFA, Rosenzweig1991PWFA, Rosenzweig1987PWFA,Blumenfeld2007EnergyDoubling,Kallos2008ExternalInjection,Litos2014ExtInjection} and laser-driven \cite{TajimaDawson1979LWFA,Pukhov2002Bubble,ManglesNature2004,FaureNature2004,GeddesNature2004,Faure2006CollidingLaser} plasma wakefield  accelerators, 
transient charge separation of plasma electrons and ions can provide ultra-strong accelerating and focusing electric fields, whose properties can be controlled by the plasma density $n_e$. 
For example, the on-axis accelerating peak electric field $E_x$ in the limits of classical wave-breaking  \cite{Dawson1959WaveBreaking, Akhiezer1956WaveBreaking} scales with the plasma density as $E_x \propto n_e^{1/2}$ just as the plasma frequency $\omega_p$; this wakefield can reach tens to hundreds of GV/m amplitude at plasma densities between $n_e \approx 10^{23}-10^{24}\,\mathrm{m^{-3}}$. As such field levels are orders of magnitude stronger than those in metallic accelerator cavities, plasma accelerators do not only represent an alternative to the unsustainably growing footprint of conventional particle accelerators  \cite{Panofsky1997evolution}, but also offer generation of ultra-high quality electron beams \cite{Hidding2012Trojan, Deng2019Trojan, Li2013IonisationInjection, Yu2014TwoColor} since the rapid acceleration limits space charge-based growth of emittance. As plasma-produced electron beams are also ultra-short in duration, down to the fs-level, no further beam compression is required. Such compression is necessary in conventional accelerators and may strongly increase the emittance, e.g. due to  coherent synchrotron radiation. The potential of plasma-based electron sources and accelerators with high initial and preserved beam quality therefore fuels a wide range  of prospective applications, including compact light sources based on free-electron lasers, inverse Compton scattering and betatron radiation \cite{Corde2013,Hidding2014,Habib2019}. Further applications extend to unique strong field and high energy physics scenarios \cite{adli2013collidersnowmass,Adli:2661806alegro2019}. 

The injection of electrons into the plasma wave is a crucial challenge, however, as this process  determines the obtainable electron beam quality. Accordingly, beam injection has represented an intensely researched topic  since the conception of plasma wakefield accelerators. The key goals of this effort are high beam quality, tunability, reliability and stability. %, as well as spurious dark current free operation.  
Various injection concepts have been developed for both laser-driven wakefield accelerators (LWFA) as well as particle-beam-driven plasma wakefield accelerators (PWFA). These include the use of colliding laser pulses \cite{Faure2006CollidingLaser, Esarey1997CollidingLP},
the generation of additional electrons via ionization of available plasma components \cite{Oz2007IonisationInjection, Umstadter1996CollidingPulses, Chen2006LaserIonisationInj, Hidding2012Trojan,Najafabadi2014IonisationInjection, Deng2019Trojan}, and the generation of plasma density downramps. The latter relies on tailoring the plasma density profile directly -- the underlying medium which provides the accelerating and focusing fields -- such that a precise subset of plasma electrons enter the accelerating phase of the wakefield in order to be captured.

Both for LWFA and PWFA, gentle  \cite{Bulanov1998Downramp} and steeper  \cite{Suk2001Downramp} density downramps have been proposed to achieve controlled injection.
On these density transitions, the phase velocity of the wakefield reduces as $v_\mathrm{ph}=c(1+\frac{1}{2}\frac{1}{n_\mathrm{e}(x)}\frac{\partial n_\mathrm{e}(x)}{\partial x}\xi )^{-1}$ \cite{PhysRevEFubianiDensityGradientsshort} at a position $\xi=x-ct$ behind the  driver  
in the co-moving frame, with $n_\mathrm{e}(x)$ being the longitudinal electron density distribution in the laboratory frame.  Consequently, the density downramp alters trajectories of ambient plasma electrons, and thereby warps and elongates the wakefield structure. This can facilitate injection of plasma electrons, and the spatial distribution of $n_\mathrm{e}$ defines the injection rate together with the resulting electron beam phase space.

Density downramp injection has been demonstrated for LWFA, where the density gradient can be generated by plasma expansion \cite{Chien2005Downramp, Faure2010Downramp, Brijesh2012Downramp}, gas flow \cite{Geddes2008Downramp, Gonsalves2011Downramp, Hansson2015Downramp}, or shock fronts in gas jets \cite{Schmid2010Downramp, Buck2013Downramp, Barber2017Downramp, Swanson2017Downramp, Burza2013Downramp}. While downramp injection is an experimentally established method in LWFA, and experimental evidence suggests it can provide even better emittance than e.g. LWFA ionization injection methods \cite{Barber2017Downramp}, successful experimental realization of downramp injection in the dephasing-free PWFA has not been achieved until very recently \cite{Deng2019Trojan}. This is despite that a large fraction of seminal downramp injection theory work was delivered in context of PWFA \cite{Suk2001Downramp,EnglandPhysRevEDensityTrans2002},  its potential as high-brightness electron beam source \cite{Thompson2004Downramp} had been discovered, and  many further theoretical and simulation-based studies with gentle \cite{XuPRABdownramp2017,JoshiFacet-II_2018,ZhangPRABfacet-iiDownramp, GREBENYUKdownrampNIM2014-2014246} and steeper \cite{DeLaOssa2017Downramp} ramps have since been carried through.

\begin{figure}%[h]
\includegraphics[width=0.5\textwidth]{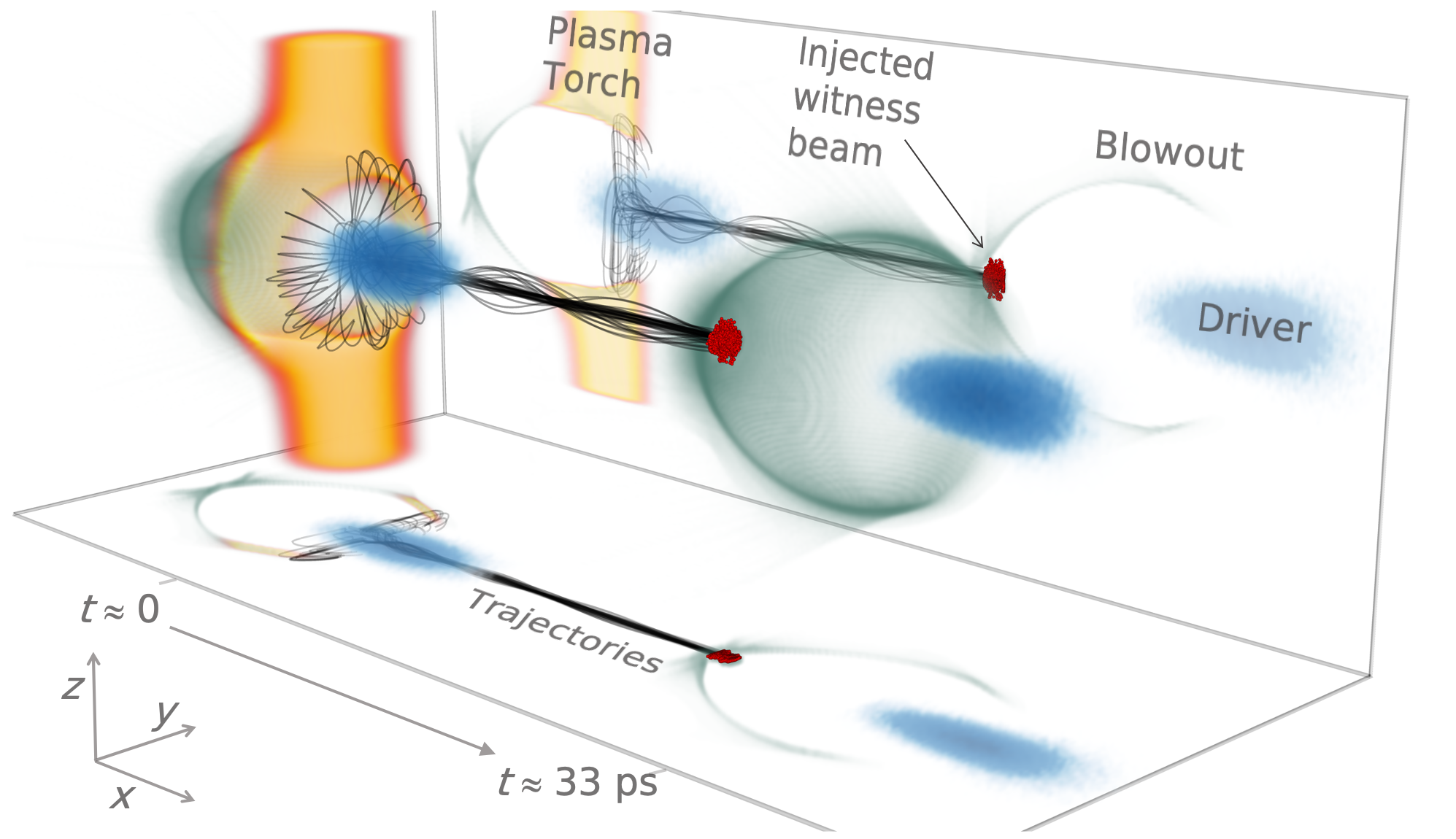}
\caption{Representation of a plasma torch injector based on 3D particle-in-cell simulations. The electron drive beam (blue) excites a plasma wave (green). At $t \approx 0$ ps and $x \approx 0$ m in the laboratory frame, this ensemble traverses the pre-formed plasma density spike -- the plasma torch (orange). This warps the plasma wave, which leads to injection of plasma electrons. Those subsequently form the witness beam (red) shown on the right hand side in a later snapshot at $t \approx 33$ ps. Selected trajectories of trapped electrons are displayed in the laboratory frame (black lines). See also \cite{supplMaterial} for the corresponding video.}
\label{fig:setup}
\end{figure}

Further theoretical work suggests improved control over injection in density downramp schemes by utilizing additional magnetic fields \cite{Zhao_2018MagneticFieldInjectionIOP}, and downramps can be used to facilitate trapping for plasma photocathodes in low-current plasma wakefield accelerators \cite{alex2014downrampassisted}. 
Furthermore, plasma density ramps are crucial elements for external injection, extraction and staging of plasma accelerators to address the challenges in beam quality preservation \cite{ASSMANN1998544transverse,anticitransport2012,MehrlingPhysRevSTAB.15.111303-2012,MiglioratiPhysRevSTAB.16.011302-2013,DornmairPhysRevSTAB.18.041302-2015,XuPhysRevLett.116.124801phasespace2016}. Tailored plasma slabs can also act as plasma lenses and (re-)focus electron beams from plasma-accelerators  \cite{KuschelLensPhysRevAccelBeams.19.071301,ActivePlasmaLensPRLTilborg2015, thaury2015Plasmalens} and linear accelerators (linacs)  \cite{PlasmaLensRosenzweigPRD1989,LitosThinPlasmaLensPRAB2019}.

In order to realize plasma density downramp injection in PWFA, we have developed the so called 'plasma torch' approach \cite{Wittig2015Downramp, Wittig2016Downramp}. The scheme exploits the -- compared to LWFA -- rather modest peak electric fields of the PWFA drive beam by adding a low-intensity laser pulse. This pulse provides higher electric fields than the driver beam, which allows production of well-defined, tunable and 'cold' plasma density regions via tunnelling ionization from ambient atoms or ions, that are otherwise unaffected by the PWFA process. 
This all-optically generated plasma torch offers flexible tailoring of the associated plasma spike and downramp distribution. In contrast to hydrodynamic approaches, which rely on re-arranging gas or plasma volumes, this method locally adds an extra plasma component $n_{\mathrm{T}}$ decoupled from the medium sustaining the PWFA.

\Cref{fig:setup} visualizes the plasma torch scheme based on a 3D particle-in-cell (PIC) simulation using VSim/VORPAL \cite{Nieter2004VSim} (further see \cite{supplMaterial}). The electron drive beam (blue) propagates from left to right through plasma, e.g. generated from low-ionization threshold gas such as hydrogen, thereby exciting an intense trailing plasma wave (green) in the blowout regime \cite{Rosenzweig1991PWFA}. 
The laser pulse generating the plasma torch density spike (orange),  e.g. from high-ionization threshold media such as helium, has already crossed the electron beam propagation axis and left the simulation box. 
The snapshot on the left at $t \approx 0$ shows the moment when the electron drive beam traverses the torch, i.e. shortly before the plasma wave leaves this volume of enhanced density and snaps back to the regular plasma  configuration. 
In this process,  ambient plasma electrons enter the blowout from specific locations within the plasma torch region as indicated by selected trajectories of to-be-captured electrons (black). 
Those -- depicted in the laboratory frame -- subsequently undergo multiple betatron oscillation periods while forming  the injected electron beam and during the acceleration process.
The snapshot on the right visualizes the situation at a later time $t \approx$ 33 ps, where this beam (red) 'witnesses' the accelerating and focusing wakefields and gains energy. 
%ere, it gains energy linearly with time until the PWFA process ceases.

In the following, we report on the first experimental realization of this all-optical plasma density downramp injection scheme, and explore its further potential and tunability with theory and simulations.\\

\section{II. EXPERIMENTAL DEMONSTRATION OF DOWNRAMP INJECTION AT SLAC FACET}
\begin{figure*}%[h]
\includegraphics[width=1.0\textwidth]{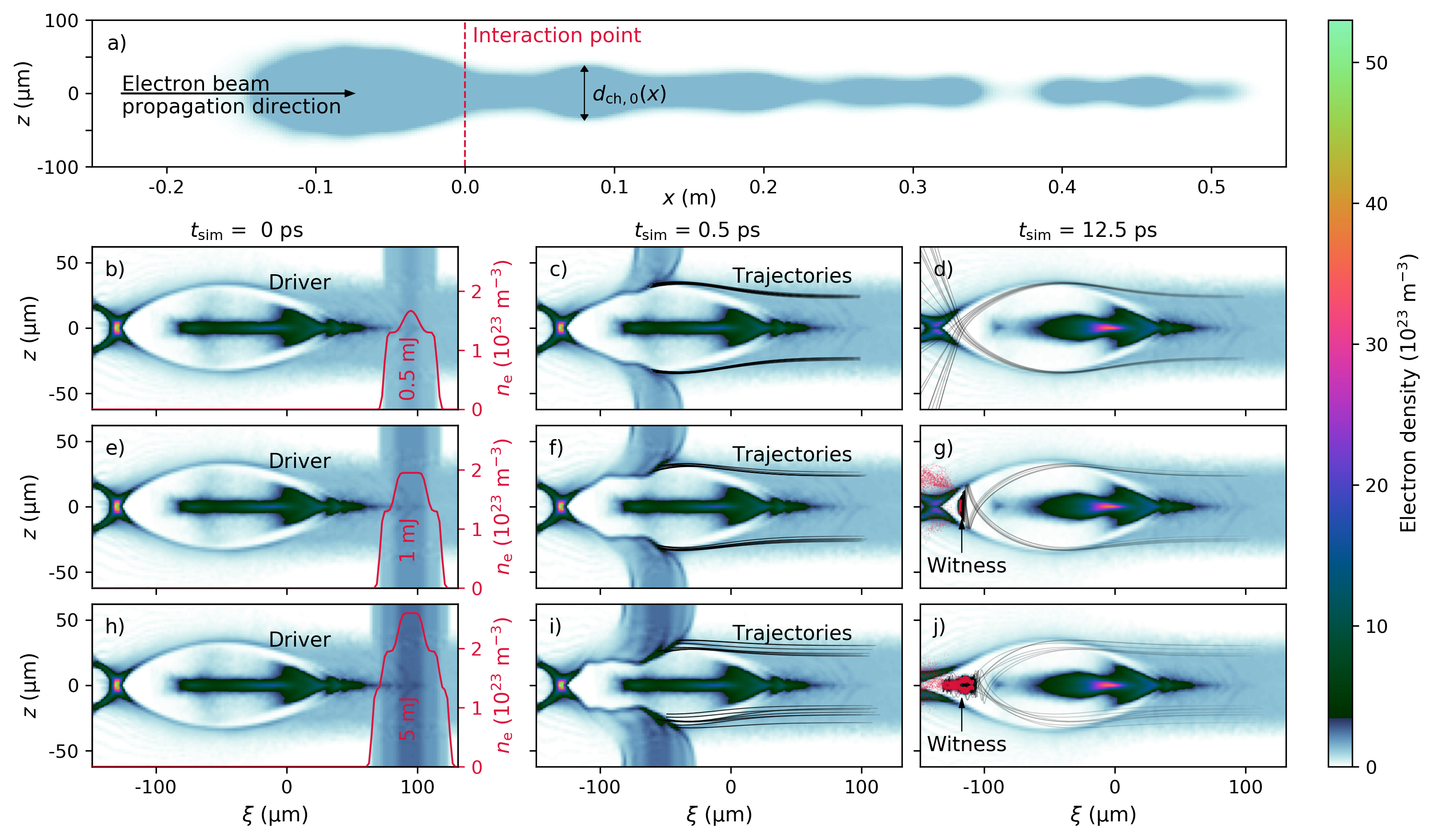}
\caption{Modeling of the FACET experiment in PIC simulations. a) shows the calculated density profile of the plasma channel generated by the axilens-focused preionization laser.  The drive beam (black) propagates to the right through the plasma (color-coded) and excites a plasma wave. Interactions with different torch distributions outlined by red lines are displayed in b) to j). Simulation snapshots for 0.5, 1 and 5 mJ torch laser energy (first, second and third row) are shown at different time steps $t_{\mathrm{sim}}$ = 0, 0.5 and 12.5 ps (left, center, right column), respectively. The plasma torch modulates the blowout structure (center column) and triggers sheath crossing of electrons as shown by selected trajectories in the co-moving reference frame (black lines). For torch laser energies $E_{\mathrm{L}} =$ 1 mJ and 5 mJ, $Q_{\mathrm{1 mJ}}{\sim} 94$ pC and $Q_{\mathrm{5 mJ}} {\sim} 498$ pC are trapped (i.e. they  exceed 5 MeV) and form the witness beam (red dots) shown in g) and j), respectively.}
\label{fig:experimentalSIMs}
\end{figure*}

We developed  capabilities required to explore and demonstrate plasma torch density downramp injection
at  the Facility for Advanced Accelerator Experimental Tests (FACET) at the SLAC National Accelerator Laboratory within the  E-210 collaboration. The linac provided electron drive beams with charge $Q_{\mathrm{D}} \approx 3.2$ nC, energy $W_{\mathrm{D}} \approx 20$ GeV, length $\sigma_{\mathrm{rms}, x} \approx  25-40\, \mathrm{\mu m}$ and typical widths of $\sigma_{\mathrm{rms}, y} \approx 15-30\, \mathrm{\mu m}$ and   $\sigma_{\mathrm{rms},  z} \approx 20-30\, \mathrm{\mu m}$, respectively. This drive beam was focused into an  experimental chamber filled with a pre-mixed 50/50 hydrogen/helium gas mixture at  ${\sim} 5.2$ mbar. The hydrogen component is fully ionized by a laser pulse focused by a holographic axilens \cite{Davidson1991Axilens,  Green2014AxilensFACET} that produces a  plasma channel of ${\sim} 1$ m length and maximum diameter of ${\sim}100\ \mu$m \cite{Deng2019Trojan}, that varies substantially in width as shown in \cref{fig:experimentalSIMs} a). Similarly tailored preionization setups were also exploited in other  plasma wakefield acceleration experiments at FACET \cite{CordeNature2015,gessner2016demonstration, Deng2019Trojan}. 

The plasma size was limited by the available spatial footprint and laser energy budget. This constrained the choice of the plasma density as the channel had to fully enclose the blowout for a sufficiently long acceleration distance. Under these circumstances, the optimum condition was chosen by employing a hydrogen plasma channel density of $n_{\mathrm{ch}} \approx 1.3 \times 10^{23}\, \mathrm{m^{-3}}$. 
The drive beam thus extended substantially into the accelerating phase of the blowout, i.e. $k_\mathrm{p} \sigma_{\mathrm{rms}, x} > 2^{1/2}$, where $k_\mathrm{p} = ( n_{\mathrm{ch}} e^2/\epsilon_0 m_e c^2)^{1/2}$ represents the plasma wave number. Here,  $e$  denotes the elementary charge, $\epsilon_0$ the vacuum permittivity  and  $m_e$ the electron mass. 

In addition, a separate laser arm was split off from the main laser path and individually compressed to a FWHM duration of $\tau_{\mathrm{L}} \approx 64$ fs. This pulse was focused perpendicularly to the electron beam path with an off-axis in-vacuum parabola (f/22.9) to a spot size of $w_\mathrm{0,L}\approx 20\, \mathrm{\mu m}$ r.m.s. at the interaction point $x = 0$ in the laboratory frame. An attenuator allowed adjusting the torch laser energy up to the maximum energy $E_{\mathrm{L}} \approx 5.1$ mJ,  corresponding to intensity levels up to  $I_{\mathrm{L}} \approx 1.2 \times 10^{16} \, \mathrm{W/cm^2}$.
Motorization of the focusing optics allowed for shifting the laser focus position along the laser propagation axis $z$ and for rotations in the $y z$-plane. This motorization facilitated versatile positioning of the plasma torch relative to the electron beam axis and the hydrogen plasma channel. At the same time, varying the laser energy changed the corresponding intensity profile and associated tunneling ionization rates \cite{nikishov1967ADK,perelomov1967ADK,Perelomov1966ADK,Nikishov1966ADK, Ammosov1986ADK,Bruhwiler2003ADK}, which altered the volume and shape of the plasma torch density distribution $n_{\mathrm{T}}$.
Generally, larger $E_{\mathrm{L}}$ can produce steeper -- and steplike -- plasma density gradients due to the generation of additional $\mathrm{He^+}$ and $\mathrm{He^{2+}}$ at higher intensities. In some experimental configurations, this has  fully depleted the ionization levels of helium and has generated very steep plasma density ramps. 
The relative time-of-arrival (TOA) between this torch-generating laser pulse and the electron drive beam was quantified by an electro-optic sampling  (EOS) setup \cite{Yan2000EOS,Scherkl2019} upstream of the interaction point with an accuracy of $\tau_{\mathrm{EOS}} \approx 25.8\ \pm\ 2.5$ fs. This provided time-stamping of the acquired data and quantified the shot-to-shot TOA jitter of $109\ \pm\ 12$ fs (r.m.s.) obtainable at FACET. An optical delay stage was used to vary the nominal TOA. %between torch laser pulse and linac-generated electron drive beam. 
The charge and energy distribution of the generated electron witness beams were measured with beam position monitors (BPMs) %question for FACET: how to cite here?
and an imaging spectrometer. %question for FACET: same, who to cite?

The conditions in the experiment are recaptured with 3D PIC simulations using a simulation box size of $500\ \mathrm{\mu m} \times 332\ \mathrm{\mu m} \times 332\ \mathrm{\mu m}$ in $x$, $y$ and $z$ with cubic cells extending over $2\ \mathrm{\mu m}$. 
\Cref{fig:experimentalSIMs} a) shows the shape of the background plasma (blue) implemented in the simulations, resembling the preionized plasma channel in the experiment with 8 particles per cell (PPC).
The initial simulation starts at the beginning of the plasma at $x = -0.2$ m to self-consistently model the drive beam (16 PPC) evolution until the injection point.
There, various plasma torch distributions can be implemented according to different laser configurations.

The properties of the laser pulse are of paramount importance for the torch process, as they determine the density distribution $n_\mathrm{T}(x,y,z)$ and, consequently, the  injected witness beam. In \cref{fig:experimentalSIMs} b)-j), simulations for three different  torch laser energy levels that were experimentally realized  \cite{Deng2019Trojan} are shown, namely   $E_{\mathrm{L}} \approx  0.5, 1.0$ and $5.0$ mJ. With increasing torch laser energy, both the peak plasma density of the torch as well as its radial extent grow as visualized by the red density profiles in snapshots b), e) and h) of \cref{fig:experimentalSIMs}. 

For the minimum torch laser energy case $E_{\mathrm{L}} \approx 0.5$ mJ, the central peak of the arising plasma torch shown in \cref{fig:experimentalSIMs} b) results from partial ionization of He in the center of the Gaussian laser pulse. This combination of torch width, peak density and gradient does not suffice for injecting and trapping electrons into the plasma wave: the blowout is not strongly perturbed such that trajectories cross the blowout sheath approximately at the center of the blowout. Electrons therefore do not gain sufficient forward momentum to stay within the blowout, which  is emphasized by  selected trajectories (black lines) shown in \cref{fig:experimentalSIMs} c) and d). 
These dynamics prevent  injection of charge in the experiment for the given torch laser energy. 
As a side note, such sub-threshold laser energy conversely allows isolation of the plasma photocathode regime, in which the laser pulse releases electrons directly inside the blowout \cite{Deng2019Trojan} where its electrostatic potential can accelerate them rapidly to the phase velocity of the wake. 

As shown in \cref{fig:experimentalSIMs} e), doubling the torch laser energy to $E_{\mathrm{L}} \approx  1$ mJ considerably changes the resulting plasma torch density profile overlaying the preionized hydrogen plasma. It particularly displays a fully ionized central flat top region formed from $\mathrm{He^+}$, thus  increasing the peak plasma density and providing steeper density ramps.
Consequently, the local deformation of the plasma blowout is more pronounced, and plasma electrons can be captured. The selected  trajectories of trapped electrons (black lines in \cref{fig:experimentalSIMs} f) and g)) cross the blowout sheath further to the front of the blowout, such that these electrons gain more energy from the elongating wakefield than in the previous case. The trajectories also indicate the region of origin of trapped electrons, which will be investigated in detail later. In this configuration, the formed beam consists of $Q_{\mathrm{1 mJ}}$ $\approx$ 94 pC trapped charge. At ${\sim}12.5$ ps after the torch location, the  kinetic energy of trapped electrons has reached $W_{\mathrm{1 mJ}} \approx$ 105 MeV (\cref{fig:experimentalSIMs} f)), corresponding to an average accelerating field of 28 GV$/$m along the %varying 
plasma channel. 

Further increased torch laser energy $E_{\mathrm{L}} \approx$ 5 mJ provides intensities up to $I_{\mathrm{L}} \approx 1.2 \times   10^{16} \, \mathrm{W/cm^2}$, which also ionizes the  second atomic level of helium and thus increases the peak plasma density to $n_{\mathrm{T}} + n_{\mathrm{ch}} \approx 2.6 \times 10^{23}\, \mathrm{m^{-3}}$. The resulting torch distribution is much wider and provides steeper density gradients, which intensifies the deformation of the plasma blowout as shown in \cref{fig:experimentalSIMs} i). This configuration injects a large amount of charge $Q_{\mathrm{5 mJ}}$ $\approx$ 498 pC as represented in \cref{fig:experimentalSIMs} j).  After injection, the blowout structure is significantly lengthened due to beam loading \cite{Wilks1987BeamLoading,TzoufrasBeamLoadingPRL2008, Manahan2017Dechirper}, which also manifests as reduced peak energy of the injected electrons of $W_{\mathrm{5 mJ}} \approx 75$ MeV in \cref{fig:experimentalSIMs} j) when compared to the 1 mJ case and \cref{fig:experimentalSIMs} g). 

These simulation results reproduce the experimental measurements in \cite{Deng2019Trojan} and demonstrate that witness beam charge can be tuned over 100's of pC simply by varying the torch laser energy. 
Control of the laser pulse parameters and exploitation of selective ionization of various ionization thresholds in hydrogen and helium or other gas mixtures thus allows designing the plasma torch profile and the resulting  witness beam properties in a wide parameter space. 

Another means of designing the injector is varying the relative time-of-arrival between the torch laser and the electron drive beam. This offers capabilities unique to all-optical configurations: in the experiment at FACET, the plasma torch could be generated by the laser pulse clearly before (TOA $\ll$ -0.15 ps) or after drive beam arrival (TOA $\gg$ 0.9 ps). For coinciding laser and electron beams at approximately ${\sim}$ 0.0 ps, the plasma torch was only partially formed when the drive beam and the trailing plasma wave arrived at the interaction point. 
Thus, the TOA varies the extent of the plasma torch available for deforming the blowout. %The longitudinal density gradients, however, remain unchanged by different TOAs as opposed to hydrodynamic schemes, which rely on temporal expansion processes. 
Systematic variation of the time-of-arrival reveals further details of the plasma torch injection process and the resulting impact for the production of witness beams.

\begin{figure}%[h]
\includegraphics[width=0.45\textwidth]{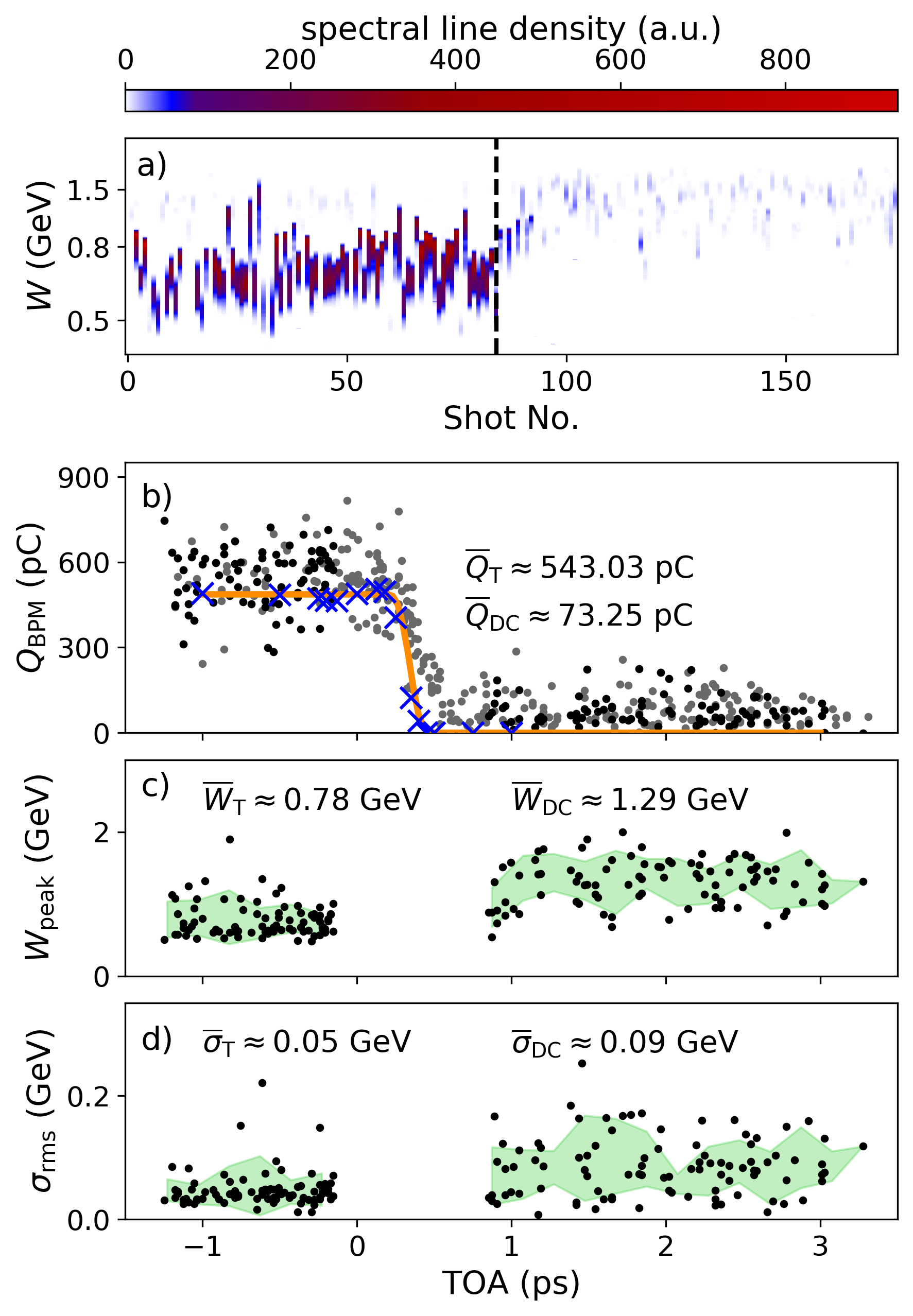}
\caption{Electron witness beams observed at FACET. a), integrated electron line spectra for laser-early  (left, TOA $<$-0.15 ps) and laser-late  mode (right, TOA $>$ 0.9). b) to d) shows witness beam properties  as function of TOA, where  TOA $=$ 0  is defined in PIC simulations (blue crosses) as coincidence of  electron beam and torch laser pulse centers. b), BPM  charge measurements (gray dots: all shots, black: shots fully resolved on the spectrometer, orange: phenomenological model in \cref{eq:overlap}). c) and d) show peak energies and r.m.s. width for shots in a), along with their standard deviation (green).}
\label{fig:timing}
\end{figure}

Electron spectrometer and BPMs were used in conjunction with the EOS to record TOA scans over a 5 ps wide time window.  \Cref{fig:timing} shows a consecutive TOA scan % for overlapping spatial alignment of plasma torch laser and electron driver beam axis 
for the experimental case of $E_{\mathrm{L}} \approx$ 5 mJ.
In \cref{fig:timing} a) on the left hand side, energy spectra of injected electron beams for the case where the laser arrives earlier than the electron beam are given, whereas on the right hand side the laser arrives after the plasma wave  and all measured injected charge is considered dark current.  While such dark current should generally be avoided for clear, well-localized and tunable electron beam generation \cite{Li2013IonisationInjection,Manahan2016HotSpots}, it can arise from beam or wakefield hot spots  \cite{Oz2007IonisationInjection, Li2013IonisationInjection,OssaIonizationPRL2013,Manahan2016HotSpots} and subsequent trapping of electrons liberated from higher ionization levels of ambient media. Here, such hot spots are promoted by wide plasma channel regions and drive beam envelope oscillations and can produce dark current \cite{Deng2019Trojan}. As quantified and shown in \cref{fig:timing} b), the average charge of this dark current amounts to ${\sim} 73 \pm 52$ pC, whereas the average charge of the plasma torch-injected electron population on top of the dark current amounts to ${\sim} 543 \pm 97$ pC. 
The rate of dark current production is influenced by shot-to-shot jitter variations of the drive beam on the one hand, and preionized plasma channel composition on the other hand, but is independent of the plasma torch laser pulses and the associated jitters. Analysis of the dark current therefore allows monitoring of the impact of these major sources of fluctuations as a subset of the contributions relevant for plasma torch injection. This may become usable to fine-tune the PWFA process in the future, in particular when torch injection becomes independent of jitters related to the plasma channel generation (see below). %The relative jitter for plasma torch produced beams is thus smaller and renders this mechanism more stable for applications. However, charge from dark current superimposes torch shots as can be seen in  \cref{fig:timing} a), meaning that its variation contributes to the torch mode jitter. 

We conduct PIC simulations analogous to \cref{fig:experimentalSIMs} h)-j) and include the torch laser in the envelope approximation to reproduce the dynamics of the plasma torch production. The witness beam charges (cf. \cref{fig:timing} b), blue crosses) obtained from simulation reproduce the transition region $-0.15$ ps $<$ TOA $<$ 0.9 ps and both plateaus observed experimentally, and further provide the absolute TOA offset for the experimental data.

The charge plateau in the plasma torch mode (negative timings, laser early) represents an injection regime practically independent of timing jitter: once generated, the cold torch density distribution does not evolve significantly over picosecond timescales, but rather over nanoseconds \cite{Durfee1993PlasmaExpansion,Shalloo2018}. This resilience to shot-to-shot TOA jitter stabilizes the production of witness beams and relaxes requirements for synchronization of drive beam and torch laser. The remaining observed shot-to-shot jitter in measured witness beam charge and energy distribution is attributed primarily to limited parameter and pointing stability of both laser pulses and the electron drive beam relative to each other. 
%The remaining observed shot-to-shot jitter is attributed to non-temporal fluctuations of the preionization and plasma torch laser pulses and the electron drive beam, in particular with regard to their spatial arrangement among one another as discussed further below. 

In the dark current mode, the measured electron spectra display energies slightly larger than those  produced by the well-localized plasma torch. It may therefore be deduced that the dark current most likely forms close to or shortly upstream of the plasma torch laser position (compare \cite{Deng2019Trojan}, suppl. Fig. 2 therein). Consequently, the beams from both modes are subject to similar wakefield dynamics downstream of the interaction point and experience the associated jitter sources.
We thus compare the obtained spectra for the dark current (laser late) and plasma torch (laser early) modes. As shown in \cref{fig:timing} a), c) and d), the average peak energies $\overline{W}_{\mathrm{DC}}\approx$   1.29 $\pm$ 0.32 GeV of the dark current shots are larger than those of the plasma torch shots $\overline{W}_{\mathrm{T}} \approx$  0.78 $\pm$ 0.23 GeV. This behavior can be explained by earlier injection and/or trapping at a different phase of the wake and/or  beam loading, which is consistent with simulations (cf. \cref{fig:experimentalSIMs} j)). It is important to note  that the width of the increasingly narrowing plasma channel along the driver beam propagation axis (see \cref{fig:experimentalSIMs} a)) limits the overall electron energy levels obtained along the PWFA  \cite{Deng2019Trojan}.

A related picture arises for the r.m.s. spectral widths. Here, the dark current mode produces spectra of width $\overline{\sigma}_{\mathrm{DC}} \approx 0.09\ \pm\ 0.05$ GeV in contrast to the narrower spectra produced by torch injection with $\overline{\sigma}_{\mathrm{T}} \approx 0.05\ \pm\ 0.03$ GeV.
Again, the observed energy spread can be a consequence of beam loading. Another contribution to the larger absolute spread of energies for dark current shots can result from its less localized production as opposed to well-localized plasma torch injection. 

The injected charge exhibits a steep linearly varying region around TOA ${\sim} 0.25$ ps that connects the plasma torch and dark current modes consistent with PIC-simulations. 
Within this narrow timing window, the laser generates plasma as the blowout passes, such that most electrons are released directly inside the wakefield. This regime is thus dominated by ionization injection \cite{Deng2019Trojan}, which gradually transforms into torch injection for TOA $\leq 0$ ps and will be further discussed in section III.

From the measured and simulated transition in \cref{fig:timing} b) we deduce that the injected witness charge is  proportional to the intersecting volume of the accelerating wakefield and the plasma torch. %, i.e. their geometric overlap.
We observe that the injected charge scales with this geometric overlap as
\begin{align}
    Q_{\mathrm{M}} &= - \frac{\pi^2 \epsilon_0}{\lambda_{\mathrm{p}}}\sqrt{\frac{n_{\mathrm{T},0}}{n_\mathrm{ch}}} \left[ \iiint  E_{\xi}  \frac{n_{\mathrm{T}}}{n_{\mathrm{ch}}} \mathrm{d} \xi \mathrm{d} y \mathrm{d}z \right]_{\mathrm{min} \, t}
    \label{eq:overlap}
\end{align}
where $E_{\xi}(\xi, y,z)$ denotes the (quasi-static) accelerating wakefield before interacting with the torch, and $\lambda_{\mathrm{p}} \approx 2 \pi / k_{\mathrm{p}}$ describes the plasma wavelength. $n_{\mathrm{T}}(\xi-ct, y,z, t)$ represents the plasma torch density distribution with peak density $n_{\mathrm{T},0}$. Here, $t$ is the interaction time which describes the motion of the plasma spike through the wakefield and any temporal dependency of $n_{\mathrm{T}}$ such as its generation by the laser pulse. The integral in \cref{eq:overlap} exhibits a convolution behavior and its minimum with respect to $t$ is proportional to the injected charge.
$E_{\xi}(\xi, y,z)$ can be obtained either from models such as \cite{Lu2006PRL,Golovanov2016Model} or PIC simulations. %$Q_{\mathrm{M}}$ corresponds to the maximum overlap of the accelerating field with the plasma torch distribution. 

In this work, we model $n_{\mathrm{T}} (\xi,y,z,t)$ based on tunneling ionization %ADK 
 calculations determined by the  torch laser intensity profile. Consequently, the local and momentary torch density also depends on the TOA of the laser pulse. 
Using $E_{\xi}$ obtained from PIC simulations for the experimental situation in \cref{fig:experimentalSIMs}, we evaluate  \cref{eq:overlap} across the TOAs in \cref{fig:timing}. 
The result of this phenomenological model is shown in \cref{fig:timing} b) by the orange line. This simple geometric approach reproduces the two plateaus as well as the transition region. It provides a direct, alternative view on the injection process compared to the established description of phase velocity retardation as an indirect consequence of the density profile. Furthermore, it correlates the 3D distributions of the wakefield with the density spike and may be applied to efficiently find a specific torch density profile for a target witness beam charge.\newline

\begin{figure}%[h]
\includegraphics[width=0.5\textwidth]{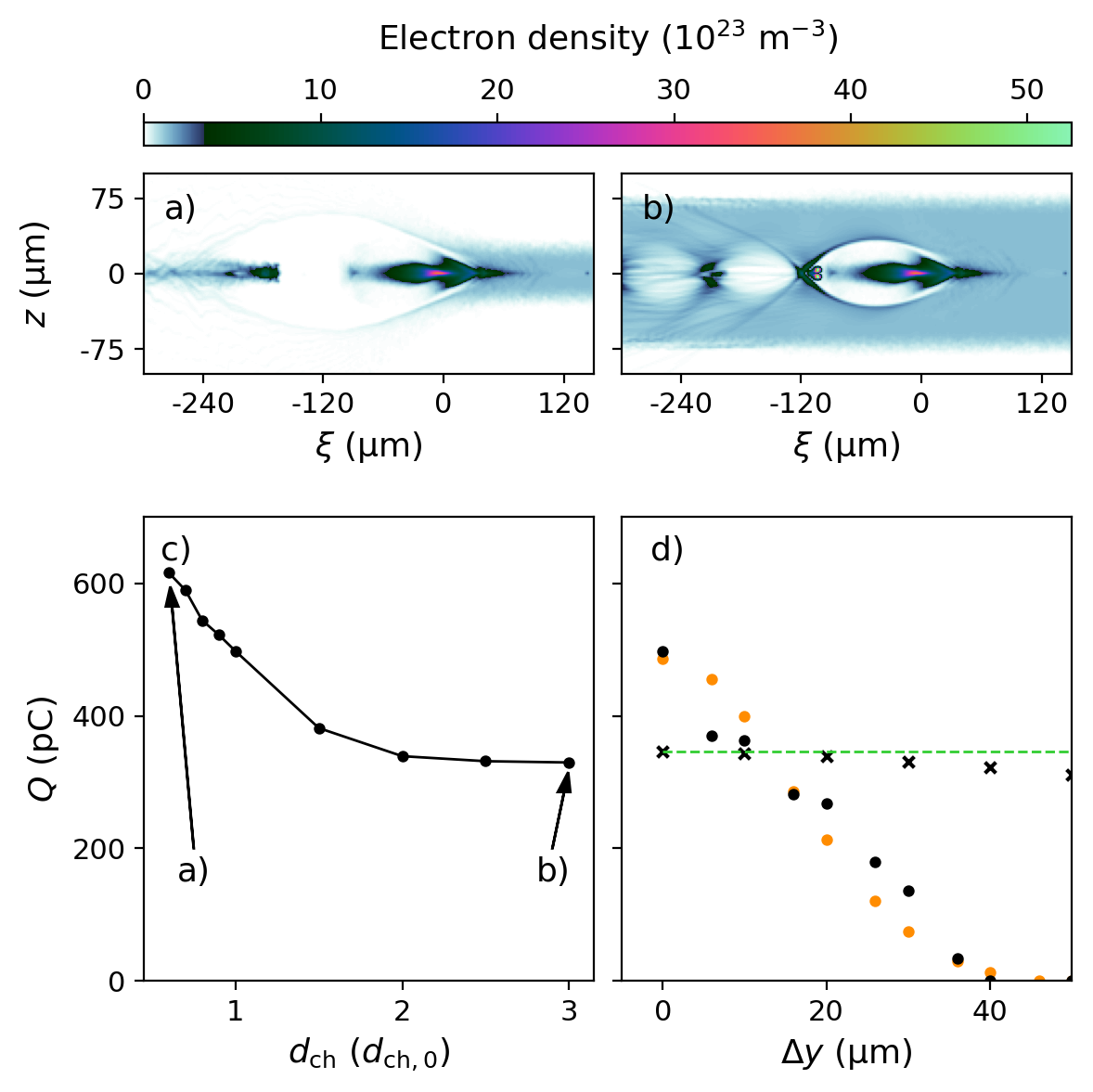}
\caption{PIC studies modeling jitter sources present in the FACET experiment. a) and b) depict snapshots of blowout formations containing torch-injected witness beams for reduced  ($d_{\mathrm{ch}} = 0.6 d_{\mathrm{ch}, 0}$) and increased channel width ($d_{\mathrm{ch}}= 3 d_{\mathrm{ch}, 0}$).  $d_{\mathrm{ch}, 0}(x)$ denotes the experimental baseline case shown in \cref{fig:experimentalSIMs}. c) shows the injected witness charge $Q$ as function of the channel width $d_{\mathrm{ch}} (d_{\mathrm{ch}, 0})$. d) depicts the simulated trapped charge $Q$ for varying torch laser misalignment $\Delta y$ for the experimental configuration (black dots) compared with the charge  $Q_{\mathrm{M}}$ obtained from \cref{eq:overlap} (orange). Further shown is the trapped charge obtained from a wider torch with flat top radius $r_{\mathrm{flat}} \approx 100\, \mathrm{\mu m} $ and wide channel ($d_{\mathrm{ch}} = 3 d_{\mathrm{ch}, 0}$) (black crosses) and the expected behavior of a wide slab-shaped torch (dashed green line).
}
\label{fig:misal}
\end{figure}

As discussed earlier, plasma torch injection can be decoupled from  TOA  jitter when operating in the distinct laser-early mode. Then, major contributions to the measured output charge jitter originate from the distribution of the preionized plasma channel, its position relative to the electron drive beam, and shot-to-shot fluctuations of the corresponding preionization laser pulse \cite{Deng2019Trojan}. In addition, the radial extent  $d_{\mathrm{ch}, 0}(x)$ of the plasma channel was  periodically narrowing along the acceleration section as shown in \cref{fig:experimentalSIMs} a). This compromises the wake excitation and renders the PWFA susceptible to various jitter sources. 
Particularly the width of the plasma channel per shot is highly sensitive to fluctuations of the preionization laser pulse parameters, which impacts on the injection yield significantly. To explore the effects of this jitter source on the injection process, we model different transverse channel width distributions  $d_{\mathrm{ch}}(x) = \kappa \times  d_{\mathrm{ch}, 0}(x)$ based on the experimental baseline case $d_{\mathrm{ch}, 0}(x)$ (cf. \cref{fig:experimentalSIMs}) in PIC simulations. $\kappa$ is varied from 0.6 to 3.0, where the former corresponds to a narrower channel as shown in \cref{fig:misal} a) and the latter resembles a much wider channel as illustrated in \cref{fig:misal} b). 
For $\kappa < 1$ scenarios,  electrons from the hydrogen plasma channel are expelled transversely beyond the stationary ion channel volume. This non-standard PWFA regime changes the re-attracting plasma forces, elongates and widens the blowout and consequently varies the electric field structure \cite{Manahan2016HotSpots}. 

Our simulations reveal two different  effects arising from  narrow channel configurations with $\kappa\leq1$. One is relevant for the plasma torch injection process itself, and the other one for the subsequent evolution of the blowout throughout the acceleration process. 
The former can be seen in \cref{fig:experimentalSIMs} for $\kappa = 1$, where trapped electrons originate from regions close to the plasma channel edges. Here, the torch laser can ionize neutral hydrogen in addition to helium. 
The resulting longitudinal density gradient $\partial n_e/\partial x$ is therefore largest \emph{outside} and at the edges of  the  thin plasma channel, where the density drops from maximum plasma torch density to zero. Inside the preionized channel, in contrast, the plasma density decreases along the electron driver beam propagation  direction only from the peak torch density to the hydrogen plasma background density. Here, the longitudinal density gradient is thus much softer than outside the channel. This transverse modulation of the longitudinal torch gradient can therefore increase the trapped charge, which further amplifies for even narrower channel widths. 
In addition, the widened blowout caused by the narrow channel changes the injection process. 
The combination of these effects impact on injected charge levels as shown in \cref{fig:misal} c). Counter-intuitively, for thinner channels, the injected witness beam charge increases  to a level of $Q \approx 616$ pC for $d_{\mathrm{ch}} = 0.6 \times d_{\mathrm{ch}, 0}$. For wider channels $d_{\mathrm{ch}} > 2 \times d_{\mathrm{ch}, 0}$, on the other hand, the blowout shrinks and gets increasingly enclosed by the plasma channel such that the trapped charge level saturates at $ 330$ pC. A further widened channel does not change  the injected charge, as the regular, uncompromised blowout size is reached. 

The second consequence from narrow and varying channel configurations affects the wakefield evolution downstream of the plasma torch and has been discussed extensively in \cite{Deng2019Trojan}. After injection, the subsequent acceleration phase depends on the evolution of the plasma channel width  along the plasma wave propagation axis $x$. The blowout size, structure and associated wakefields change when the local plasma channel width narrows below the regular blowout size, which can be approximated by the plasma wavelength $\lambda_{\mathrm{p}} \approx 100\,\mathrm{\mu m}$.  Consequently, injected witness beams experience various accelerating and focusing fields depending on local $d_{\mathrm{ch}} (x)$. The experimentally observed output energies of witness beams thus fluctuate with jittering channel generation as reflected by the range of spectra shown in \cref{fig:timing} a). We conclude that for thin plasma channels with $d_{\mathrm{ch}}$ smaller than the unperturbed blowout radius, stable witness beam generation and acceleration requires precise control over the  preionization laser profile for the plasma channel around the electron beam propagation axis. 
Alternatively and preferably, using channel radii wider than encountered in the experimental proof-of-concept situation at FACET, e.g. $d_{\mathrm{ch}}$ $\gg$ $d_{\mathrm{ch}, 0}$,  can effectively resolve these adverse influences. The setup then converges to the ideal textbook PWFA configuration, which provides a blowout structure \textit{independent} of shot-to-shot plasma channel variations. Wider channels with uniform density profile thus stabilize the injection process against the aforementioned peculiarities caused by transversally varying longitudinal density gradients e.g. at the edge between the channel and the torch filament, and they also stabilize the subsequent acceleration process.

Another source for experimentally observed  witness beam fluctuations are transverse shot-to-shot fluctuations of the torch laser propagation axis e.g. as result of limited  pointing stability. In combination with a finite size of the produced plasma torch, this entails variation of longitudinal density gradients $\partial n_e/\partial x$ across the blowout diameter.  
To study this effect, we perform simulation scans with plasma torches shifted relative to the electron drive beam propagation axis by $\Delta y$. Complementary to the partially generated, but spatially centered plasma torch occurring in the TOA transition region (see \cref{fig:timing} b)), wave-breaking and injection for an off-centered torch only occurs in the reduced overlap volume of wakefield and plasma torch as indicated by \cref{eq:overlap}. For the experimental channel width  $d_{\mathrm{ch}} = d_\mathrm{ch,0}$, this spatial asymmetry reduces the injected witness charge with increasing misalignment as can be seen in \cref{fig:misal} d) (black dots). A similar trend results from the phenomenological model (orange dots).
%Why linearly? Is this expected from the model?
This effect can vary the injected witness beam charge over hundreds of pC, consistent with experimental observations as shown in \cref{fig:timing}. While misalignment of the torch laser can be detrimental to injection stability especially in case of thin plasma channels and smaller blowouts, asymmetric injection on the other hand offers the possibility to deliberately produce asymmetric witness beams and to steer betatron oscillations as discussed later. 

Any relative spatial shot-to-shot jitter between torch laser and driver beam propagation axis can be eliminated in a similar fashion as for the plasma channel: by increasing the transverse extent of the plasma torch until it exceeds the blowout diameter significantly. To show this and rule out any effect arising from limited channel width, we  increase the latter to $d_{\mathrm{ch}} = 3 d_{\mathrm{ch}, 0}$ in another $\Delta y$-scan. Further, the plasma torch flat top radius (cf. Appendix A) is changed to $r_{\mathrm{flat}} \approx 100\, \mathrm{\mu m} $ and now fully covers the unperturbed blowout. This improves the resilience of the injector against torch laser misalignment substantially, as shown in \cref{fig:misal} d) (black crosses). For example, a misalignment of 20 $ \mathrm{\mu m}$ (${\sim}33\ \%$ of the blowout diameter)  reduces the injected witness charge by only ${\sim} 6.5$ pC or ${\sim} 2\ \%$ compared to central alignment. The reason for this slight but non-vanishing deviation is the curvature of the plasma torch. Implementing a torch shaped as planar plasma slab (cf. Appendix A) of sufficient width can completely remove this effect. Then, shot-to-shot relative alignment jitter does not impact the injected charge  at all, as indicated by the green dashed line in \cref{fig:misal} d). To achieve this immunity, the plasma slab has to be wider than the combined extent of blowout radius and maximum spatial shot-to-shot  jitter of the torch laser relative to the electron beam propagation axis.

In conclusion, the generation of plasma density spikes completely covering the blowout region  by a sufficient margin  makes plasma torch injection and acceleration resilient against shot-to-shot jitter of the preionization laser. This includes the width of the channel as well as its positional arrangement relative to the electron driver beam axis. In addition, shot-to-shot spatial jitter of the plasma torch laser relative to the wakefield propagation axis can be eliminated by shaping the plasma torch into a sufficiently wide volume or slab. These two error sources have been major limitations for the output parameter stability in the proof-of-concept experiments at FACET. 
If, on the other hand, the spatial extent of plasma channel and torch together with their (relative) alignment can be sufficiently controlled shot-by-shot, exotic configurations such as thin plasma channels for non-standard PWFA modes and asymmetric plasma torch injection can be harnessed. 
These findings, and solutions, now guide the  design for upcoming setups and plasma torch installations.\newline

\section{III. SIMULATION-BASED EXPLORATION OF THE PLASMA TORCH PARAMETER SPACE}
After establishing the findings from the experimental study, we now investigate plasma torch injection without limitations of the plasma channel width or spatial jitter. The wide plasma medium therefore now fully contains the PWFA, and increased plasma density $n_{\mathrm{ch}}$ = $6 \times 10^{23}$ $\mathrm{m^{-3}}$ corresponding to  $\lambda_{\mathrm{p}} \approx 40\,\mathrm{\mu m}$ allows for higher accelerating gradients and the exploration of benefits for the witness beam quality \cite{Thompson2004Downramp}. 
The preionized channel consists of fully ionized hydrogen and the first level of helium, $\mathrm{He^+}$, to mitigate dark current from field ionization (see Appendix B). The plasma torch density spike results from ionizing helium a second time, thus yielding $\mathrm{He^{2+}}$. For the purpose of this study, we omit detailed investigations of the required laser intensity distributions. Instead, we study the capabilities of the plasma torch method more fundamentally by exploring the effect of different three-dimensional plasma torch density profiles in a systematic manner. 

In the following simulations and as outlined in Appendix A, various plasma torches are modeled as cylinders extending in $z$-direction with a central flat top radius $r_\mathrm{flat}$ corresponding to full ionization at peak densities $n_\mathrm{T, 0}$ superimposing the plasma channel. Cosine-shaped ramps of total length $l_\mathrm{ramp}$ connect the flat top density and the background plasma density $n_{\mathrm{ch}}$ in radial direction ($r^2 = x^2 + y^2$) around the torch laser propagation axis. Similar profiles are obtained for the laser-generated distributions in \cref{fig:experimentalSIMs}, which describe the FACET experiments.

The simulations employ electron drive beam parameters obtainable at SLACs FACET-II facility \cite{FACET-IIPhysRevAccelBeams.22.101301Yakimenko2019}, tuned for dark current free PWFA \cite{Manahan2016HotSpots} for the given plasma density. For instance, the Gaussian driver beam contains charge of $Q_{\mathrm{D}}$ = 0.6 nC within a length $\sigma_{\mathrm{rms}, z}$ = 7.5 $\mathrm{\mu m}$ and has an energy $W_{\mathrm{D}}$ = 10 GeV corresponding to a Lorentz factor $\gamma_{\mathrm{D}} \approx 2\times10^5$ and energy spread $\Delta W/W_{\mathrm{D}}$ = 0.02, and normalized emittance $\epsilon_{\mathrm{rms, n}}$ = 100 mm mrad in both planes.  To avoid envelope oscillations of the drive beam, the transverse size is matched to $\sigma_{\mathrm{rms}, x} = \sigma_{\mathrm{rms}, y} = ( \epsilon_{\mathrm{rms, n}} \lambda_{\mathrm{p}} / 2 \pi \sqrt{2/\gamma_{\mathrm{D}}} )^{1/2} \approx 3\, \mathrm{\mu m}$.  The maximum blowout radius observed from simulations is $r_{\mathrm{b}} \approx 15\, \mathrm{\mu m}$. These simulations employ a moving window and cell sizes of 0.25 $\mathrm{\mu m}$  and 0.5 $\mathrm{\mu m}$ in longitudinal and transverse directions, respectively, as well as a modified Yee solver \cite{Cowan2013}. The electron beam consists of 16 PPC and the plasma is modeled by 8 PPC (16 PPC in the plasma torch region).  

\begin{figure}%[h]
 \includegraphics[width=0.5\textwidth]{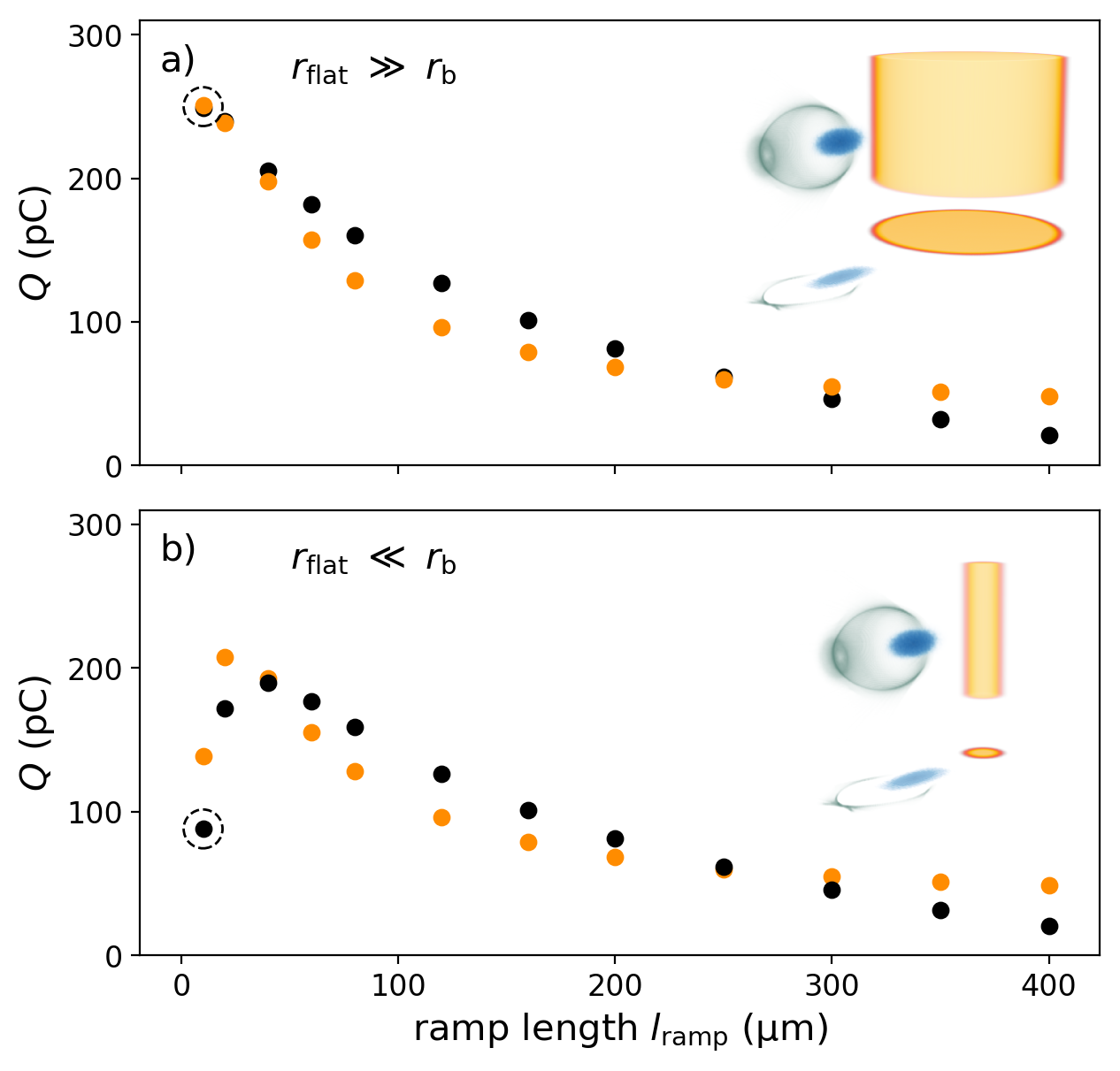}   
\caption{Injected charge $Q$ in dependence of torch geometry.  a), wide plasma torches with $r_{\mathrm{flat}}$ $\gg$ $r_{\mathrm{b}}$ and b), narrow plasma torches with $r_{\mathrm{flat}}$ $\ll$ $r_{\mathrm{b}}$ for different ramp lengths. Black: PIC-simulations, orange: $Q_{\mathrm{M}}$ modeled via \cref{eq:overlap}. The insets visualize the short ramp scenarios for the cases highlighted by the dashed circles.}
\label{fig:rampScan}
\end{figure}

\Cref{fig:rampScan} presents results from injection studies with fixed peak torch density $n_{\mathrm{T}, 0} = 0.5 \times n_{\mathrm{ch}}$ superimposing the plasma channel. At this torch density, no witness beam generates dark current via tunnelling ionization (see Appendix B). The following simulations vary the torch ramp length $l_{\mathrm{ramp}}$ for two different cases: torch density columns with i) wider and ii) narrower radial flat top extent than the blowout radius. 
The first case, shown in \Cref{fig:rampScan} a), covers plasma torches with flat top radius $r_{\mathrm{flat}} = 40 \, \mathrm{\mu m} \gg r_{\mathrm{b}}$. In contrast to our experiments at FACET where the torch diameter was barely as large as the blowout diameter, this allows wave breaking across the full cross section of the blowout. The cylindrical radius of curvature of this torch distribution is large compared to the plasma wave's transverse extent, such that the injection process is approximately symmetric in the $y z$-plane. 
We scan ramp lengths from $l_{\mathrm{ramp}} = 10 \ \mathrm{\mu m}$ -- highlighted by the dashed circle and the 3D-inset in \Cref{fig:rampScan} a) -- up to $l_{\mathrm{ramp}} = 400 \, \mathrm{\mu m}$,  and present the trapped witness beam charge (black dots) along with values obtained from the phenomenological model (orange dots, cf. \cref{eq:overlap}). 
This wide scan range comprises different regimes of injection: with long ramps  $l_{\mathrm{ramp}} > \lambda_{\mathrm{p}}$, injection occurs predominantly from the rear of the blowout, while with short ramps $l_{\mathrm{ramp}} < \lambda_{\mathrm{p}}$ the blowout collapses off-axis closer to the drive beam and sudden re-phasing provides injection as in \cref{fig:experimentalSIMs}.
These different injection mechanisms are explored in theory ever since \cite{Bulanov1998Downramp} (gentle ramps, in context of LWFA) and \cite{Suk2001Downramp} (steep ramps, in context of classical hydrodynamic downramp-based PWFA) and have fueled different approaches for optimization of injected beam quality for hydrodynamic downramp injection for PWFA and LWFA  \cite{BrantovPoP2008DensityInhomogeneity,XuPRABdownramp2017, DeLaOssa2017Downramp,Ekerfelt2017Downramp}. 

We repeat the same scan for wide, uniform plasma slabs approximating idealized conventional density downramp configurations and find similar injection dynamics, along with almost identical amounts of trapped charge as for the wide plasma torch. For both geometries, more charge is trapped for steep ramps, and less for softer ramps in agreement with \cite{Suk2006Patent}. 
We thus infer that wide plasma torches can inherently mimic  conventional density downramp configurations with both steep and gentle ramps as a subset of their range of capabilities.

As wide torches with short ramps feature maximal injected charge for a given plateau density, only steep gradients provided by the plasma torch scheme may facilitate injection in PWFA driven by comparatively low-current electron beams. Further amplification can be achieved by increasing the plateau density as discussed in Appendix B.  These strategies enable PWFA applications in accelerators lacking the intense electron beams available at FACET.

Additionally, the plasma torch technique facilitates unique plasma density spikes thinner than those in our proof-of-concept experiments described in Section II.  
In fact, the flat top radius can be  significantly smaller than the blowout radius. In a corresponding simulation scan shown in \cref{fig:rampScan} b), it is set to $r_{\mathrm{flat}} = 2.5 \,\mathrm{\mu m} \ll r_{\mathrm{b}}$. Increasing the torch ramp length in this mode of operation changes the radial torch extent substantially, and thus effectively enhances the overlap volume between torch and blowout in $x$ and $y$ until the  blowout interacts with the plasma spike similarly to a wide torch. 
The trapped witness beam charges therefore converge to those associated with wide torches as in \cref{fig:rampScan} a) for ramps   $l_{\mathrm{ramp}} \gtrsim  100 \, \mathrm{\mu m}$. 
Reducing the ramp length, on the other hand, increases the trapped charge similarly to the steep ramps in the wide-torch case. However, this scenario exhibits a global maximum of injected charge at $l_{\mathrm{ramp}} \approx 50 \, \mathrm{\mu m}$,  formerly not observed in density downramp schemes. 
In this range, the typical impact of shorter ramps  -- namely injecting ever higher charge -- is  overpowered by the narrowness of the created plasma torch profile. Then, $r_{\mathrm{flat}} + l_{\mathrm{ramp}} \lesssim r_{\mathrm{b}}$ and the  plasma torch only covers the full extent of the blowout along the main torch axis in $z$, i.e. in the propagation direction of the plasma torch-generating laser pulse. In the perpendicular radial direction $y$, the narrow torch density profile reduces the overlap volume with the blowout. Wake deformation and wave breaking therefore happen asymmetrically, only affect a subset of the blowout and thus change the injection process substantially. Even for this regime, the phenomenological model through \cref{eq:overlap} reproduces the overall distribution of $Q(l_{\mathrm{ramp}})$ obtained from simulations. %Deviations occurring for small $l_{\mathrm{ramp}}$  indicate the influence of further dynamics.

Both ramp scans shown in  \cref{fig:rampScan} highlight the capability of plasma torch PWFA to realize a wide range of different density downramp physics and seamlessly switch between them, e.g. by changing the intensity profile of the torch laser. 
In contrast to typical downramp approaches, however, plasma torch injectors prove highly versatile as their density spike freely superimposes the PWFA medium.
For example, they can be located arbitrarily along the PWFA and even multiple instances can be created in short succession. A double-torch injector is presented Appendix C and demonstrates generation of multi-color beams, e.g. applicable in pump-probe experiments.

In addition to freely locate plasma torch injectors, we show in the following sections that shaping the plasma torch density profile $n_{\mathrm{T}}(x,y,z,t)$ in space and time facilitates control not only over injected charge, but also over the resulting witness beam distribution and quality. The initial density distribution of trapped plasma electrons is  of critical importance, as it represents the initial conditions for the injection process and thus governs the properties and evolution of the witness beam. By back-tracking all trapped plasma electrons in PIC-simulations, we reconstruct and visualize this \emph{trapping volume}.\newline

\begin{figure*}%[h]
 \includegraphics[width=1.0\textwidth]{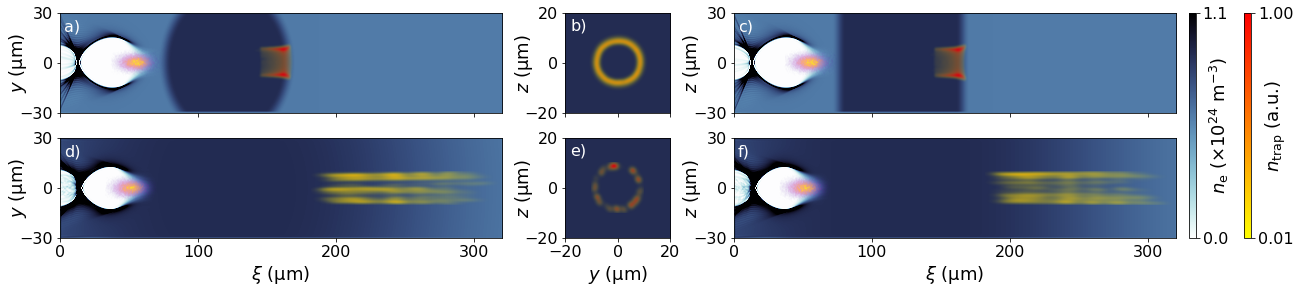}
\caption{Visualization of trapping doughnuts obtained from PIC-simulations in the plane transverse to the axial dimension of the torch (left column), along electron drive beam axis  (center column) and torch side view (right column). The two rows correspond to \cref{fig:rampScan} a), the wide torch, with short ramp $l_{\mathrm{ramp}} = 10\ \mathrm{\mu m}$,  and long ramp  $l_{\mathrm{ramp}} = 200\ \mathrm{\mu m}$, respectively.}
\label{fig:donutsSymm}
\end{figure*}

\subsection{SYMMETRIC INJECTION}
As shown in the previous sections, wide torches exhibit similar features as conventional downramp schemes. 
This particularly results from radially symmetric interactions between the blowout and the plasma torch, which implies a symmetric drive beam and isotropic torch density distribution across the blowout extent.  
\Cref{fig:donutsSymm} visualizes simulations where the electron drive beam (orange) propagates to the right through the plasma represented as color-coded slice through the simulation box center. 
The plasma torches shown here combine the plateau region with radius $r_{\mathrm{flat}} = 40 \, \mathrm{\mu m}$ with short $l_\mathrm{ramp} = 10 \,\mathrm{\mu m}$ and long ramps $l_\mathrm{ramp} = 200 \,\mathrm{\mu m}$, respectively. These cases were already shown in \cref{fig:rampScan} a). 
The projected trapping volumes $n_{\mathrm{trap}}(\xi,y,z)$ associated to the subsequently trapped witness beams are color-coded in yellow-red.  
The first column in \cref{fig:donutsSymm} shows the projection in plasma torch laser propagation direction ($x y$-plane), the center column shows the projection in the propagation direction of the electron beam driver ($y z$-plane), and the third column shows the side-view of the plasma torch ($x z$-plane). These densities are normalized across each column of the figure. 
For the given interaction, the longitudinal projection of the trapping volume (central column) displays a characteristic annular shape similar to \cite{DoppLight2017,DeLaOssa2017Downramp} around the drive beam propagation axis. We observe this effect for all interactions between radially symmetric blowouts and wide plasma torches and call this fundamental structure \textit{trapping doughnut}.

The longitudinal extent of the trapping volume of the wide torch with short ramp length $l_\mathrm{ramp} = 10 \,\mathrm{\mu m}$ in \cref{fig:donutsSymm} a)-c) slightly exceeds the downramp length. It broadens radially towards the downstream end of the ramp (see \cref{fig:donutsSymm} a) and c)) due to the expanding blowout on the density downramp. The trapping doughnut as depicted in \cref{fig:donutsSymm} b) in head-on view is perfectly symmetric  and regular due to the likewise symmetric interaction.
As already shown in \Cref{fig:rampScan} a), the corresponding injected and trapped charge amounts to $256$ pC and agrees with the phenomenological model \cref{eq:overlap} $Q_{\mathrm{M}} \approx 251$ pC. 

\begin{figure}%[h]
 \includegraphics[width=0.49\textwidth]{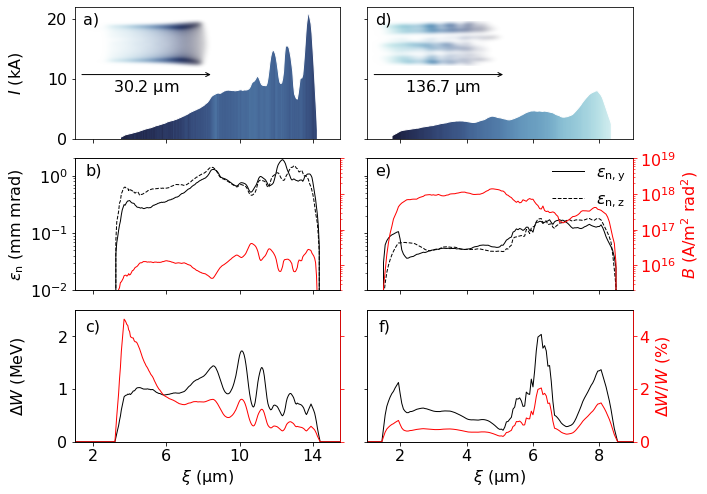}
\caption{Comparison of witness beam slice parameters injected by wide plasma torches ($r_{\mathrm{flat}}$ = 40 $\mathrm{\mu m})$ with short ($l_{\mathrm{ramp}} = 10\ \mathrm{\mu m}$) a)-c) and long ($l_{\mathrm{ramp}} = 200\ \mathrm{\mu m}$) d)-f) ramps.
a) and d) show current profiles of respective witness beams, color-mapped to the longitudinal position within the trapping volume (insets). Slice normalized emittances in both planes (black) and brightness (red) are shown in b) and e). Slice energy spreads are given in c) and f).
}
\label{fig:sliceParameters}
\end{figure}

The trapping volume for the wide torch with long ramp $l_\mathrm{ramp} = 200 \,\mathrm{\mu m}$ remains circular, but extends over much longer distance than for short ramps. 
In contrast to the trapping volume in the previous case, however, the long ramp generates an irregular density pattern consisting of multiple injection filaments. 
We attribute this inhomogeneity to the sensitivity of the adiabatic injection process, as the soft gradient reduces the efficacy of the density ramp. 
Similar irregularities were reported for LWFA self-injection \cite{DoppLight2017}, outlining comparable injection dynamics as for long  downramps in PWFA. 
Trapping is therefore susceptible to inhomogeneities of the driver beam, wakefield and plasma profile -- and is computationally prone to noise and resolution \cite{Silva2020Downramp}. 
Repeating the simulation at refined grid resolution of 0.1 $\mathrm{\mu m}$ in the longitudinal direction resembles a similarly irregular trapping pattern (cf Appendix D).
The witness beam generated on this gentle ramp has a charge $82$ pC ($Q_{\mathrm{M}} \approx 68 $ pC) and exhibits slices with particularly low  emittance as shown in \cref{fig:sliceParameters}, consistent with findings in \cite{XuPRABdownramp2017}.   

As a side note,  the annular trapping doughnut shape remains similar for a case with equally long ramp but shorter, 2.5 $\mathrm{\mu}$m-long  radius (see Appendix D), since $l_{\mathrm{ramp}} + r_{\mathrm{flat}} \gg r_{\mathrm{b}}$ is still fulfilled. However, the injected charge reduces slightly to $80$ pC ($Q_{\mathrm{M}} \approx 68 $ pC) due to the weak torch density modulation at the edges of the blowout.

We now investigate witness bunch parameters produced by the two trapping volumes shown in \cref{fig:donutsSymm} in more detail. \Cref{fig:sliceParameters} a)-c) shows key slice parameters (50 nm bin size) of the witness beam resulting from the short ramp visualized in \cref{fig:donutsSymm} a)-c). \Cref{fig:sliceParameters} d)-f) contrasts them with the witness beam injected on the long ramp shown in \cref{fig:donutsSymm} d)-f). The short ramp case with injected charge $256$ pC (${\sim} 4.8\times 10^5$ macro particles) forms a beam with $\sigma_x \approx 2.6\, \mathrm{\mu}$m r.m.s. length,  whereas the long ramp case is formed of $82$ pC (${\sim} 1.7\times10^5$  macro particles) within $\sigma_x \approx 1.8\ \mu$m r.m.s. length. The average obtained witness electron energies amount to ${\sim}{103}$ MeV and ${\sim}{106}$ MeV for the short and long-ramp case, respectively.

In \Cref{fig:sliceParameters} a) and d), the corresponding current profiles for both beams are given. The current produced by the steep ramp  exceeds the current of the gentle ramp by more than a factor of 2 and reaches peak values close to $I_\mathrm{p}\approx20$ kA. The current profiles are color-coded for each slice of the witness beam by the average longitudinal origin position within the trapping doughnut, shown as $x y$ trapping region  in the top-left inset. For the gentle ramp, this mapping reveals a strictly linear relation, expressing that electrons residing at the plateau-end of the density ramp form the head of the witness beam and vice versa. The beam generated by the steep torch gradient, in contrast, lacks a clear correlation, as electrons injected from multiple initial positions cross trajectories and produce mixed slices  within the formed witness beam.

With regard to beam quality, both torch configurations produce low projected emittances, e.g. $\epsilon_{{\mathrm{n}}} \lesssim 1.6 $ mm mrad for the steep ramp and $\epsilon_{{\mathrm{n}}} \approx 0.4 $ mm mrad for the gentle ramp. The slice emittances shown in \cref{fig:sliceParameters} b) and e) vary strongly along both beams, but are mostly below 1 mm mrad. Particularly the beam generated on the gentle ramp displays a long high-quality region where the slice emittance remains below $0.1$ mm mrad in both planes. This resembles brightness values up to $B = 2 I_\mathrm{p}/(\epsilon_{{\mathrm{n,y}}}\epsilon_{{\mathrm{n, z}}}) \approx 10^{18} \mathrm{A/(m\ rad)^2}$  -- many orders of magnitude larger than obtainable in conventional accelerator systems. 
\Cref{fig:sliceParameters} c) and f) present the absolute slice energy spreads of both beams, which are, except for a few slice positions and in particular for the gentle ramp, below 1 MeV. The latter further exhibits low slice energy spreads below $\Delta W \lesssim 0.6$ MeV  concomitant with the lowest emittance region. This renders this particular slice range  exceptionally well suited for demanding applications such as high brightness light sources.\newline

 %Left: 256 pC, right: 82

\subsection{ASYMMETRIC INJECTION}
The plasma torch process allows injection from asymmetric interactions of the plasma wave with the torch density distribution. Modification of torch profiles and the corresponding trapping doughnut can be used to produce unique witness beam modalities.

\begin{figure*}%[h]
 \includegraphics[width=1.0\textwidth]{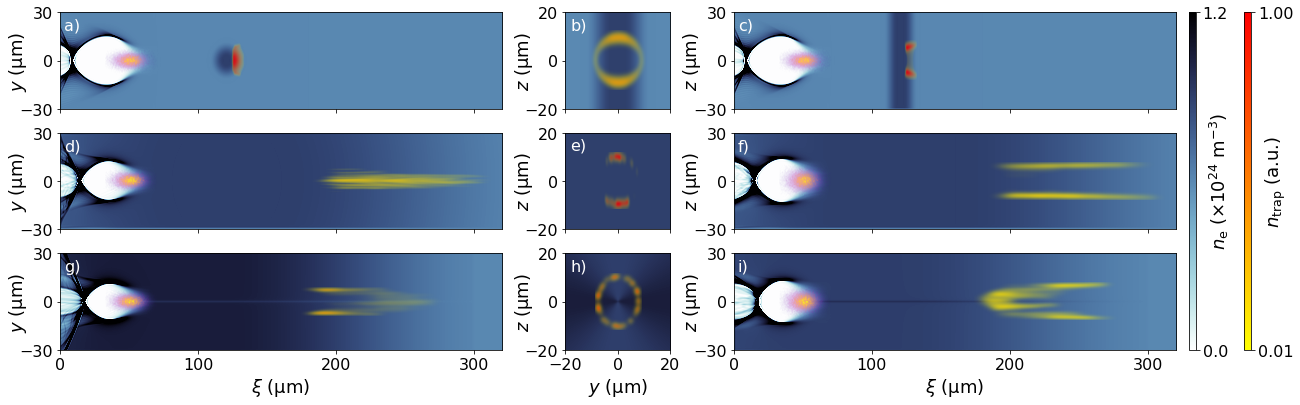}
\caption{Visualization of trapping doughnuts obtained from PIC simulations analogous to \cref{fig:donutsSymm}. The first row corresponds to the narrow torch with short ramp $l_{\mathrm{ramp}} = 10\ \mathrm{\mu m}$ shown in \cref{fig:rampScan} b). The second row displays an identical configuration as in \cref{fig:donutsSymm} d)-f) but with an asymmetric drive beam $\sigma_{\mathrm{z}} = 1.5 \times \sigma_{\mathrm{y}}$ yielding an asymmetric blowout with radius $r_{\mathrm{b},z} \approx 1.1\times r_{\mathrm{b},y}$. The third row repeats the former, extended by an additional torch density modulation $n = n_{\mathrm{T}} \times [1 + 0.5 \cos{(\phi(z,y)})^2]$.}
\label{fig:donutsAsymm}
\end{figure*}

The first case shown in \cref{fig:donutsAsymm} revisits a configuration with narrow spatial extent of the plasma torch shown in  \cref{fig:rampScan} b): $r_{\mathrm{flat}} + l_{\mathrm{ramp}} = 12.5 \,\mathrm{\mu m} < r_{\mathrm{b}}$. Here, the trapped charge reduces significantly to $88$ pC ($Q_{\mathrm{M}} \approx 139$ pC) compared to  $256$ pC ($Q_{\mathrm{M}} \approx 251$ pC) trapped on the wide torch shown in \cref{fig:donutsSymm} a)-c). 
\Cref{fig:donutsAsymm} a)-c) visualizes a substantially diminished overlap of the thin torch with the plasma blowout in different planes: as shown in \cref{fig:donutsAsymm} a) and b), the reduced transverse overlap in $y$-direction effectively crops the trapping volume and removes parts of the typical annular shape (cf. \cref{fig:donutsSymm} b). Additionally, the shorter longitudinal extent of the torch reduces the amount of plasma available for trapping.
This unique interaction geometry precipitates non-isotropic injection predominantly in the $x z$-plane. As we shall see later, this causes the formation of twin populations of injected electrons, which originate from the pronounced asymmetry along the $z$-direction of the trapping doughnut depicted in \cref{fig:donutsAsymm} c).

Next to spatially asymmetric torch density distributions, the formalism in \cref{eq:overlap} also predicts that radial asymmetries in the blowout can change the injected charge substantially. This occurs for example for transversely asymmetric drive beams. Those are the norm rather than the exception both for linacs, due to beam compression and focusing techniques, as well as for laser-plasma accelerators, where linear laser pulse polarization \cite{ManglesPRL2006PhysRevLett.96.215001,DoppLight2017,Couperus2017} produces transversely elliptical electron beam profiles. 

In \cref{fig:donutsAsymm} d)-f), a drive beam with $\sigma_\mathrm{z} = 1.5 \times \sigma_\mathrm{y}$ ($\epsilon_{\mathrm{n},y} =\epsilon_{\mathrm{n},z}$) generates an asymmetric blowout with radii  $r_z \approx 1.1 \times r_y$.
We study the injection into this wake formation with a wide torch and long ramp identical to \cref{fig:donutsSymm} d)-f) and find that such asymmetric blowouts deform the trapping doughnut substantially: 
instead of a full $360^{\circ}$ doughnut, trapping is allowed only in a highly confined angular range around the $z$-axis, as shown in \cref{fig:donutsAsymm} e). 
Remarkably, this angular selection does not result from cropping the circular trapping doughnut observed in the symmetric drive beam case  \cref{fig:donutsSymm} d)-f), but from re-arranging the original distribution. 
Injection in the $x y$-plane, where the blowout is thinner, is suppressed, while it is promoted in the $x z$-plane, where the blowout is wider. 
In fact, the total injected witness charge  \emph{increases} slightly by ${\sim} 6\ \%$ to 87 pC (increases by ${\sim} 10\ \%$ to $Q_{\mathrm{M}} \approx 76$ pC). Enhanced injection rates in the $x z$-plane thus (over-)compensate the missing parts of the trapping doughnut. %This demonstrates that  blowout asymmetries can impact the trapping doughnut substantially.
The strongly planar injection driven by an asymmetric driver beam produces two distinct witness beam populations, even more pronounced than for the cropped trapping volume in \cref{fig:donutsAsymm} a)-c) conversely arising from a thin torch and a symmetric driver beam,  and without the loss of charge.
However,  the witness beam emittances produced from the more realistic, asymmetric driver beam exceed the particularly low emittance produced for the perfectly symmetric drive beam substantially. Similar emittance increase has been observed in \cite{ZhangPRABfacet-iiDownramp} for asymmetric driver beams. 
This configuration and the analysis of resulting witness beam emittance in different planes is discussed further in \cref{fig:beamlets}.  

We also investigate  the effect of the same asymmetric drive beam on injection from a wide torch with short ramps (see Appendix D).  This variation results in a homogeneous doughnut that is, however, stretched in $z$-direction, where the  drive beam is wider, and compressed in the $y$-direction, both by approximately $20\ \%$. Compared to the symmetric driver beam shown in \cref{fig:donutsAsymm} a)-c), the trapped charge reduces by $10\ \%$ ($Q_{\mathrm{M}}$ reduces by $5\ \%$) and the emittance does not change substantially. This indicates that steep downramp injection is, in some aspects, less sensitive to drive beam asymmetries % and hence better decoupled 
than gentle downramp injection.

Since the trapping process in the gentle downramp case is sensitive to asymmetries in the wakefields, we conjecture that the asymmetry of the driver can be compensated to some extent by a suitable modulation of the plasma density. The plasma torch technique in principle allows such modulations, e.g. by employing specifically shaped, or multiple laser pulses. 
In a first exploration, the torch is modulated by a radial density function $n = n_{\mathrm{T}} \times [1 + 0.5 \cos(\phi(z,y))^2]$  with $\phi$ being the polar angle in the $z y$-plane, to partially counteract the effect of the asymmetric drive beam. 
Indeed, this can partially compensate the imbalance of trapped charge and prevent generation of beamlets: as shown in \cref{fig:donutsAsymm} g)-i), increasing the torch density in the narrow dimension of the asymmetric drive beam recovers a more symmetric trapping  doughnut (albeit elliptic, similar to those for asymmetric driver on a short ramp shown in Appendix D)  and forms a single beam with more symmetric emittance. Such plasma torch based modulation may indicate a potential path to produce small slice emittances from gentle downramp injection even in case of more realistic, namely asymmetrically shaped driver beams.

\begin{figure*}%[h]
 \includegraphics[width=1.0\textwidth]{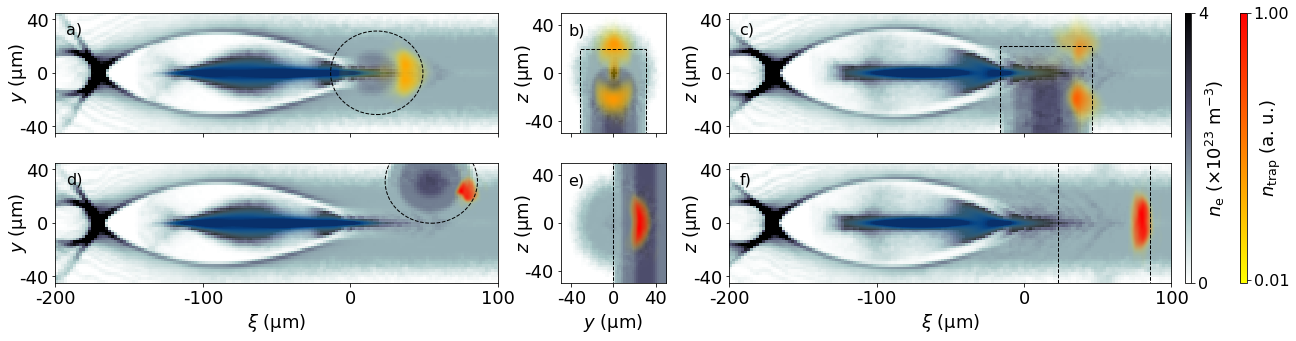}
\caption{Visualization of trapping doughnuts analogous to \cref{fig:donutsSymm} and \cref{fig:donutsAsymm} for simulations modeling experiments at FACET. The first row corresponds to the timing scan shown in \cref{fig:timing} with TOA = $-0.2$ ps. The dashed box marks the extent of the laser-generated plasma torch. Results of a torch displaced transversely by $\Delta y = 30\, \mathrm{\mu m}$ as in  \cref{fig:misal} is shown in the second row.}
\label{fig:donutsExp}
\end{figure*}

Finally, the two cases presented in \cref{fig:donutsExp} recall the experimental conditions at FACET discussed in section II. 
The scenario shown in \cref{fig:donutsExp} a)-c) involves a narrow plasma torch that is transiently generated while the blowout passes the ionization front of the laser pulse. 
This simulation corresponds to a TOA $\approx -0.2$ ps between  electron drive beam and torch laser for the timing scan summarized  in \cref{fig:timing} b). % and is located on the left of the steep transition region.  
In the PIC snapshot shown in \cref{fig:donutsExp} c), the torch laser pulse propagates in positive $z$-direction. Its corresponding ionization front, indicated by the dashed line, just traversed the driver beam axis.
Consequently, the arising plasma torch filament is only partially formed along the $z$-direction when the drive beam arrives at the interaction point. The laser pulse continues to ionize across the radial extent of the blowout during the subsequent interaction, partially inside the blowout, which leads to injection of $Q \approx$ 466 pC ($Q_{\mathrm{M}} \approx$ 486 pC).
The resulting trapping doughnut is cropped in the $y$-direction and partially resembles the case in \cref{fig:donutsAsymm} a)-c), since the torch radius is smaller than the blowout radius. This effect is overlayed by the significant asymmetry of the drive beam and the temporal dependence of the torch density spike in $z$-direction.  
Furthermore,  direct ionization injection via the plasma photocathode mechanism \cite{Deng2019Trojan} contributes charge to the witness beam. The signature of this mixed-mode injection manifests in the charge originating from the center of the trapping doughnut shown in \cref{fig:donutsExp} b), which does not occur for fully formed torches (cf. \cref{fig:donutsSymm} and \cref{fig:donutsAsymm}). 
For TOA values $> -0.2$ ps, we find that the trapping doughnut increasingly changes such that the half-moon structures fade, while the central ionization injection feature strengthens, reflecting the transition from full plasma torch mode,  across direct ionization injection towards  laser-late mode. A similar mixed mode of ionization- and self-injection has been observed for LWFA \cite{DoppLight2017}.
The PIC simulations shown in \cref{fig:donutsExp} a)-c) reveal injected beamlets oscillating in $z$-direction. This likely results from the combination of the larger drive beam extent in $z$ and the cropped trapping doughnut in $y$-direction. During the experiment at FACET, however, the experimental setup did not allow conclusive observation and evidence of beamlets, likely due to limited resolution of the electron spectrometer in the witness beam energy range, and scattering elements in the beamline. Further observation, study and exploitation of the twin beamlet formation thus requires  suitable experimental conditions and diagnostics.

The second simulation shown in \cref{fig:donutsExp} d)-f) corresponds to the experimental situation of fully evolved, but transversely offset  plasma torch configuration with respect  to the electron driver beam axis. Here, the torch is shifted in $y$-direction by $\Delta y = 30 \,\mathrm{\mu m}$ as already presented in \cref{fig:misal} d), and yields injection of $Q \approx$ 137 pC ($Q_{\mathrm{M}} \approx$ 74 pC).
This specific kind of asymmetry facilitates injection only  in regions $y > 0$ where plasma torch and blowout overlap. The resulting single half-moon structure is shown in \cref{fig:donutsExp} e). Consequently, this configuration  generates a single bunchlet that oscillates in the $xy$-plane, e.g. in the plane perpendicular to previously discussed beamlets. The similarly large transverse momentum of the whole injected population manifests in large-amplitude oscillations of the beam centroid well suited for betatron radiators. % model: 74.4 pC, PIC: 137

All presented simulations show modified trapping volumes when the spatiotemporal overlap volume between the torch density profile $n_{\mathrm{T}}(x,y,z,t)$ and the plasma wave deviates from ideal radial symmetry. These regions of origin determine the initial phase space distribution of the formed witness beam and enable the production of a wide range of electron beams with exceptional properties. 

Of those, planar injection and the controlled production of counter-oscillating beamlet twins represents a particularly interesting capability. We investigate the slice properties of produced witness beams and their emittance evolution for two different pathways to generate such beamlets. Videos presenting the corresponding real space and phase space evolution of both beams can be found in \cite{supplMaterial}.

\Cref{fig:beamlets} a)-d)  describe the  beam produced from the narrow torch with short ramp, fulfilling $r_{\mathrm{flat }} + l_{\mathrm{ramp}} < r_{\mathrm{b}}$ already shown in \cref{fig:donutsAsymm} a)-c). %88 pC?
The cropped trapping doughnut is given as inset in \cref{fig:beamlets} a), and is the origin of injection of two counter-oscillating witness beam populations  as depicted by macroparticles in the 3D real-space snapshot. This beam is extracted ${\sim}$ 0.8 mm downstream of the torch position, corresponding to an energy of ${\sim}$ 100 MeV. Electron  macroparticles colored in red originate from the upper half of the trapping doughnut, corresponding to $z > 0$, and blue ones from the lower half where $z < 0$. The transverse projection in the $x z$-plane highlights the two clearly separated beamlets. The slice current and energy spread profiles are given in \cref{fig:beamlets} b), showing a significantly reduced length of the witness beam population(s) compared to the witness beam analyzed in \cref{fig:sliceParameters}.

\Cref{fig:beamlets} c) and d) depict the evolution of the projected  emittance along the plasma accelerator of the individual beamlets as well as of the combined beam in both transverse planes. In $y$-direction, both beamlets populate a congruent transverse phase space area $y y'$ as shown in the inset of \cref{fig:beamlets} c). The full beam's emittance in this plane thus equals the emittance of individual bunchlets. In $z$-direction, on the other hand, each beamlet occupies a separate phase space area $z z'$ as shown by the inset in \cref{fig:beamlets} d). 
%, each larger than in the other plane.
The emittance of both beamlets combined hence amounts to ${\sim}\sqrt{2}$ times a single beamlet emittance in this $z$-direction. The projected emittance both of individual beamlets, as well as of the combinbed beam, is larger in the $z z'$-plane than in the $y y'$-plane as consequence of the thin torch orientation in $z$-direction.

\begin{figure*}%[h]
\includegraphics[width=1.0\textwidth]{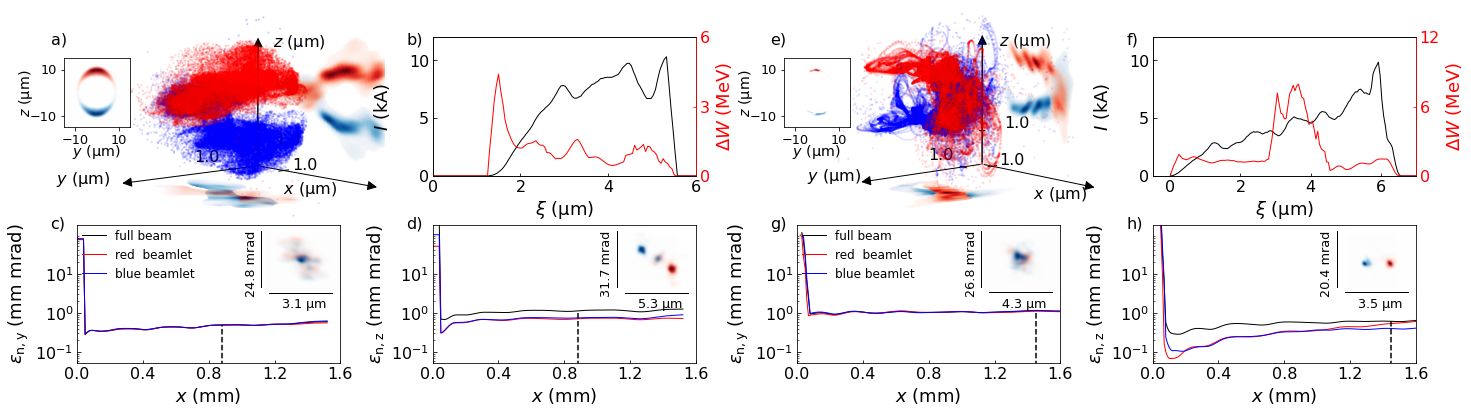}
\caption{Twin beamlet structures obtained from PIC simulations. a)-d) display the beam obtained from the thin torch with short ramp shown in \cref{fig:donutsAsymm} a)-c), and  e)-h) present the beam generated on a gentle ramp traversed by an asymmetric ($\sigma_z = 1.5\times \sigma_y$) drive beam shown in  \cref{fig:donutsAsymm} d)-f).
a) and e), 3D real space snapshots with projections, showing beamlet twins, counter-oscillating in the  $x z$-plane. b) and f), slice current and energy spreads.  c), d) and g), h)  show the evolution of the transverse witness beam emittance for the full beam and individual beamlets in both planes. The insets show transverse phase space snapshots corresponding to a) and b). The  dashed vertical line marks the labframe position where quantities were extracted from the simulations, i.e. when the beam reaches approximately 100 MeV energy. Two videos present the overall simulated evolution of real space and phase space for both beam configurations \cite{supplMaterial}.}
\label{fig:beamlets}
\end{figure*}

\Cref{fig:beamlets} e)-h) present the analysis of beamlet production from a wide torch with gentle ramps $l_{\mathrm{ramp}} > r_{\mathrm{b}}$, but asymmetric drive beam with $\sigma_{\mathrm{z}} = 1.5 \times \sigma_{\mathrm{y}}$, i.e. the  configuration shown in \cref{fig:donutsAsymm} d)-f). Here, the injection of the two distinct counter-oscillating electron populations in the $z$-direction is based on a re-arranged rather than  a cropped trapping doughnut, as discussed. \Cref{fig:beamlets} e)-f) summarize the real space, slice current and energy spread profile of the produced witness population(s). Here, the beam reaches ${\sim} 100$ MeV later at approximately $1.4$ mm (indicated by the dashed lines) downstream of the torch center because it is trapped at wakefield phases with lower amplitudes compared to the previous (thin torch, short ramp) case.
The emittance evolution and transverse phase space snapshots are shown in \cref{fig:beamlets} g)-h), respectively. 
Again, while the emittance of beamlets  as well as the combined beam is similar in $y$-direction, the combined emittance of beamlets in the plane of oscillation ($zx$) is significantly larger than the emittance of each individual beamlet.
However, in contrast to the thin torch case which crops the trapping doughnut, the beamlet emittances in the $z$-direction are substantially \emph{lower} than in the $y$-direction. The very distinct pair of beamlets is visualized by the high transverse phase space density shown in the inset in \cref{fig:beamlets} h), and manifests in projected beamlet emittance values  $\epsilon_{\mathrm{n},z} \lesssim 0.7$ mm mrad.\\

These methods of tunable, counter-oscillating twin beamlet production are highly interesting for radiation production, for example via tunable betatron radiation. In context of the $\Delta y$-scan in \cref{fig:misal} we identified the potential to eliminate one beamlet and excite planar betatron motion of a single electron population. 
As shown by the twin beamlets in \cref{fig:beamlets}, different methods can be applied to achieve planar injection both on short as well as long ramps. 
This flexibility in combination with high currents and low emittance represents excellent prospects for the generation of polarized betatron radiation  \cite{phuoc2008Betatron,DoppLight2017} with a high degree of tunability. \newline%Among other applications, these twin beamlets could also be applied to investigate unique modalities for high-field experiments.\\  

\section{SUMMARY AND CONCLUSIONS}
Plasma density downramp injection has been proposed as a method for high-quality electron beam production many years ago. 
While highly successful in laser-driven plasma accelerators, it had eluded realization in electron beam-driven plasma wakefield accelerators so far. 
We realized such long sought-after  density downramp injection -- by an advanced, all-optical, flexible generalization of downramp injection at the SLAC FACET facility. 
This first experimental proof-of-concept of the plasma torch approach demonstrates an electron beam injector controlled by a mJ-class laser pulse. 
We show that the injected beams can be tuned by the laser pulse energy, relative temporal delay and spatial alignment between plasma torch laser and plasma wave. 
Furthermore, we develop strategies towards stabilization of witness  beam generation even under unfavorable experimental conditions with regard to stability of incoming beams. Among these approaches, plasma torches wider than the blowout radius prove to generate witness beams that are particularly resilient against shot-to-shot jitter. This result offers pathways towards stable and reliable plasma-based injectors.

Based on the experimental observations, we have developed a simple, yet powerful phenomenological  model that predicts the trapped charge from torch injection. It agrees well with simulations for arbitrary spatiotemporal torch density distributions and may be combined with analytical wakefield theories to design plasma torch injectors.

Additional simulations are used to explore the further potential of the scheme and reveal the influence of the characteristic trapping volume on witness beam properties. 
It can be manipulated by various torch density distributions, including steep and soft downramps, as well as by specific blowout configurations.  
In future studies, the trapping volume, representing the initial conditions of the subsequently formed witness beam, may be exploited for precise tailoring of plasma torches that yield optimized witness beam phase space distributions. 
Particularly scenarios breaking the -- typically radial -- symmetry of downramp injection provide unique pathways to shaping the resulting witness beams, including witness beams performing large-amplitude betatron oscillations and counter-oscillating beamlet twins. 
These may be particularly interesting for generating polarized x-ray radiation. 

Plasma torch-based electron beam properties can reach charges up to hundreds of pC, tens of kA-level currents, and emittances in the sub-0.1 mm mrad range concomitant with low slice energy spreads. The corresponding slice brightness values exceed state-of-the-art of conventional accelerators by orders of magnitude. These features have promising implications for future accelerators and applications such as in photon science and high energy and high field physics. We anticipate that plasma torch based injectors -- due to their capability combined with high degree of experimental feasibility -- will be adapted and further exploited by other linac-driven as well as hybrid laser-plasma wakefield accelerators \cite{Hidding2010Hybrid, Gilljohann2019L2PWFA}.\newline

%\newpage
\section{ACKNOWLEDGEMENTS}
\begin{acknowledgments}
The FACET ‘E-210’ plasma wakefield acceleration experiment was built and operated with support from UCLA (US Department of Energy (DOE) contract no. DESC0009914), RadiaBeam Technologies (DOE contract no. DE-SC0009533), and the FACET E200 team and DOE under contract no. DE-AC02-76SF00515. 
B.H., P.S., A.S., F.A.H., T.H., A.B. were supported by the European Research Council (ERC) under the European Union’s Horizon 2020 research and innovation programme (NeXource, ERC Grant agreement No. 865877).
The work was supported by STFC ST/S006214/1 PWFA-FEL, EPSRC (grant no. EP/N028694/1).  
D.L.B. acknowledges support from the US DOE Office of High Energy Physics under award no. DE-SC0013855. J.R.C. acknowledges support from the
National Science Foundation under award no. PHY 1734281. 
M.D.L acknowledges support from the US DOE Office of High Energy Physics under award no. DE-SC0017906. 
This work used computational resources of the National Energy Research Scientific Computing Center, which is supported by DOE DE-AC02-05CH11231, and of the Supercomputing Laboratory at King Abdullah University of Science \& Technology (KAUST) in Thuwal, Saudi Arabia. \newline
\end{acknowledgments}

D.U. and P.S. contributed equally to this work.\\

\section{APPENDIX}
\appendix

\onecolumngrid
\bigskip
\section{Appendix A: Plasma torch density distributions}
The plasma torch technique allows all-optical shaping of various plasma density distributions that superimpose the plasma facilitating the PWFA. In this work, the plasma torch profile is generated either directly in the VSim PIC code, or by mapping tunneling ionization rates \cite{nikishov1967ADK,perelomov1967ADK,Perelomov1966ADK,Nikishov1966ADK,Ammosov1986ADK,Bruhwiler2003ADK} corresponding to the intensity profile of the laser externally before loading into VSim. Here, we concentrate on a subset of possible shapes, namely cylindrical and slab-like profiles. 

The cylindrical shape corresponds to the FACET E-210 experimental case and is generally valid for symmetric Gaussian laser pulses with soft focusing. We consider configurations where the core cylinder of radius $r_{\mathrm{flat}}$ along the laser propagation axis $z$ consists of fully ionized plasma.
This core is surrounded by cosine-shaped ramps of length $l_{\mathrm{ramp}}$.  The radially symmetric  distribution with $r^2 = x^2 + y^2$ reads

\begin{align}
    \frac{n_{\mathrm{T}}}{n_{\mathrm{T}, \, 0}} (r, z) =
    \begin{cases}
    1 & \text{: } r < r_{\mathrm{flat}} \\
    \cos{^2 \left( \frac{\pi}{2} \frac{r - r_{\mathrm{flat}}}{l_{\mathrm{ramp}}} 
    \right) } & 
    \text{: } r_{\mathrm{flat}} \leq r < r_{\mathrm{flat}} + l_{\mathrm{ramp}} \\
    0 & \text{: } r_{\mathrm{flat}} + l_{\mathrm{ramp}} < r.
    \end{cases}
\label{eq:torchDistro}
\end{align}

The second density distribution considered in this work is a slab. A suitable laser pulse configuration which generates such a shape could for example be a combination of two crossed cylindrical lenses. The distribution reads:  
%Slab density distribution with Cartesian coordinates:

\begin{align}
    \frac{n_{\mathrm{Slab}}}{n_{\mathrm{Slab}, \, 0}} (x,y,z) =
    \begin{cases}
    1 & \text{: } |x| < r_{\mathrm{flat}} \\
    \cos{^2 \left( \frac{\pi}{2} \frac{|x| - r_{\mathrm{flat}}}{l_{\mathrm{ramp}}} 
    \right) } & 
    \text{: } r_{\mathrm{flat}} \leq |x| < r_{\mathrm{flat}} + l_{\mathrm{ramp}} \\
    0 & \text{: } r_{\mathrm{flat}} + l_{\mathrm{ramp}} < |x|.
    \end{cases}
\label{eq:slabDistro}
\end{align}

\twocolumngrid

\bigskip
%\bigskip
%\bigskip
%\bigskip
\section{Appendix B: Plasma torch density scan}\label{appendix:B}
%Additionally, or as alternative to shaping the plasma torch distribution \cref{eq:torchDistro} or \cref{eq:slabDistro}, 
The peak torch density $n_{\mathrm{T}, \, 0}$ has a fundamental impact on the injection process. For a fixed intensity distribution of the torch laser, this changes the density gradient and the amount of plasma present in the corresponding trapping doughnuts.  
When dealing with mixtures of low ionization threshold (LIT)  and high ionization threshold (HIT) gases, the HIT component is independently tunable from the LIT component and can be adjusted simply by changing partial pressures in the gas mix reservoir. In case of the experiments at FACET, for example, a mixture of hydrogen (LIT) and helium (HIT) was used. However, at elevated LIT densities, the associated wakefield amplitude increases, and electric field hot spots at the wake vertex and/or the compressed drive beam can exceed the tunneling ionization threshold of the HIT medium \cite{Manahan2016HotSpots}, which can produce dark current. 
 
To overcome this limitation and to operate at higher LIT densities, one can switch to using a HIT medium with even higher ionization threshold. Then, much higher electric fields can be tolerated without danger of dark current production. For our simulation study in section III we thus choose the combination of fully ionized molecular hydrogen and the first helium level $\mathrm{He^+}$ as amalgamated LIT medium, forming the plasma channel density $n_{\mathrm{ch}}$. %The remaining $\mathrm{He^+}$ ions form the HIT medium for generation of density spikes. 
The $\mathrm{He^+}$ species is then the HIT medium and the transition $\mathrm{He^+} \rightarrow \mathrm{He^{2+}}$ is exploited  for generation of density spikes.  The tunability of LIT vs. HIT densities is then limited, but still accessible by varying the partial pressure ratio of hydrogen and helium in the mixture. Switching off dark current thus comes at the price of coupled LIT and HIT densities $n_{\mathrm{ch}} = n_{\mathrm{H}} + n_{\mathrm{He}}$ and $n_{\mathrm{T}} = n_{\mathrm{He}}$. This composition allows for tuning the torch density in the range $n_{\mathrm{T}}$ = $0$ to $1$ $\times n_{\mathrm{ch}}$ while maintaining the LIT density. This can be achieved by adjusting the hydrogen component $n_{\mathrm{H}}$ accordingly.

%BH: it could be comment to mention the intensiyt level required for this He2+ level, and to add an estimation of what that means for the a0 and correpsonding temperature, residual momentum etc.

\begin{figure}[h]
 \includegraphics[width=0.48\textwidth]{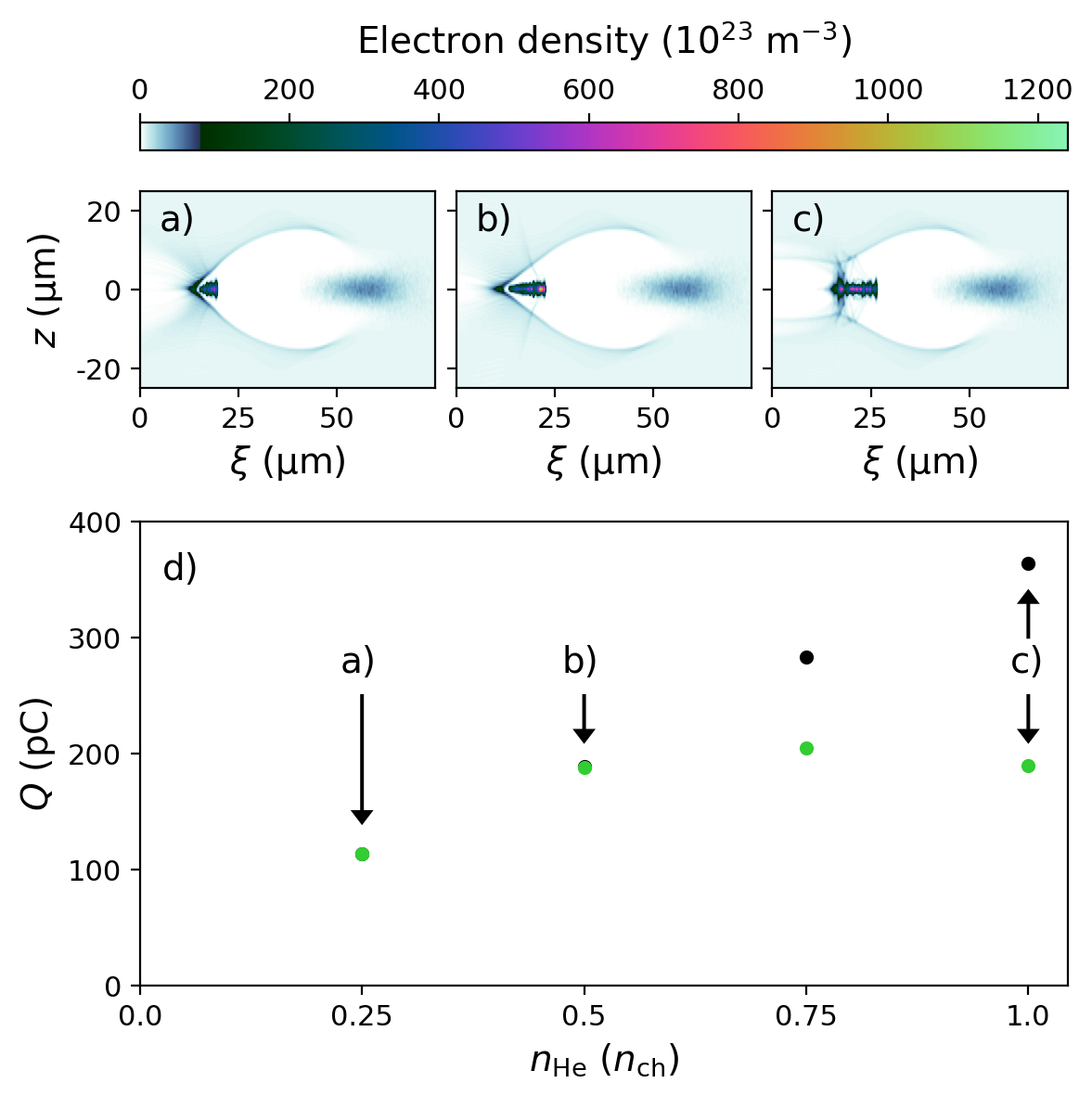}
\caption{Scan of the neutral helium density $n_\mathrm{He}$ at constant $n_\mathrm{ch}=6 \times 10^{23}\, \mathrm{m^{-3}}$. a) to c) show PIC snapshots after injection, where c) displays charge densities that lead to further witness beam induced ionization and trapping of $\mathrm{He^{2+}}$. d) shows trapped witness charge $Q$ as function of $n_\mathrm{He}$ for witness beams including (black) and excluding ionized $\mathrm{He^{2+}}$ (green).}
\label{fig:densityScan}
\end{figure}

To explore the effect of HIT density variation, we conduct simulations with constant plasma density $n_{\mathrm{ch}} = 6 \times 10^{17}\, \mathrm{cm^{-3}}$, and neutral helium densities $n_{\mathrm{He}}$ varying in a range from $0.25$ to $1 \times n_\mathrm{ch}$. The plasma torch radius $r_\mathrm{flat} = 2.5\ \mathrm{\mu m}$ and the ramp length $l_\mathrm{ramp} = 40\ \mathrm{\mu m}$ (cf. \cref{fig:rampScan} b) are kept constant. Results of this HIT density scan are presented in \cref{fig:densityScan}, where panels a)-c) show exemplary blowout structures after injection taken from simulations with $n_{\mathrm{He}}$ = $0.25$ $\times n_{\mathrm{ch}}$, $0.5\times n_{\mathrm{ch}}$ and $1.0 \times n_{\mathrm{ch}}$, and d) shows injected charge as function of $n_{\mathrm{He}}$. Increasing $n_{\mathrm{He}}$ corresponds to higher peak plasma torch densities and steeper density gradients, which consequently yields larger injected witness beam charges. This is reflected by increasing beam-loading and increasingly deformed blowout structures. While the torch-based injected charge reaches a maximum when $n_{\mathrm{He}} > 0.5 \times n_{\mathrm{ch}}$ (\cref{fig:densityScan} d), green), an additional injection mechanism sets in. Then, the fields of the torch-generated witness beam begin to exceed the threshold for tunnel ionization of the remaining $\mathrm{He^+}$ ions. This accumulates a second witness beam component \cite{Amorim2019Inception} that is added to the initial torch-injected beam. \Cref{fig:densityScan} d) (black) shows the full witness beam charge, which, after onset of witness beam self-ionization, linearly increases with $n_{\mathrm{He}}$. This represents another method for production of high-charge, multi-color beams in addition to \cite{Najafabadi2019Multicolor, Zeng2015Multicolor}.

%The capability of flexible localization of the plasma torch simply by steering the laser pulse, its efficiency in producing suitable plasma torches already at sub-mJ laser energies, and its shapeabilty also allows for multiple tunable injectors at arbitrary positions $x$ along the accelerator stage. 

\bigskip
\section{Appendix C: Multiple beams from multiple torches}

\begin{figure}%[h]
 \includegraphics[width=0.48\textwidth]{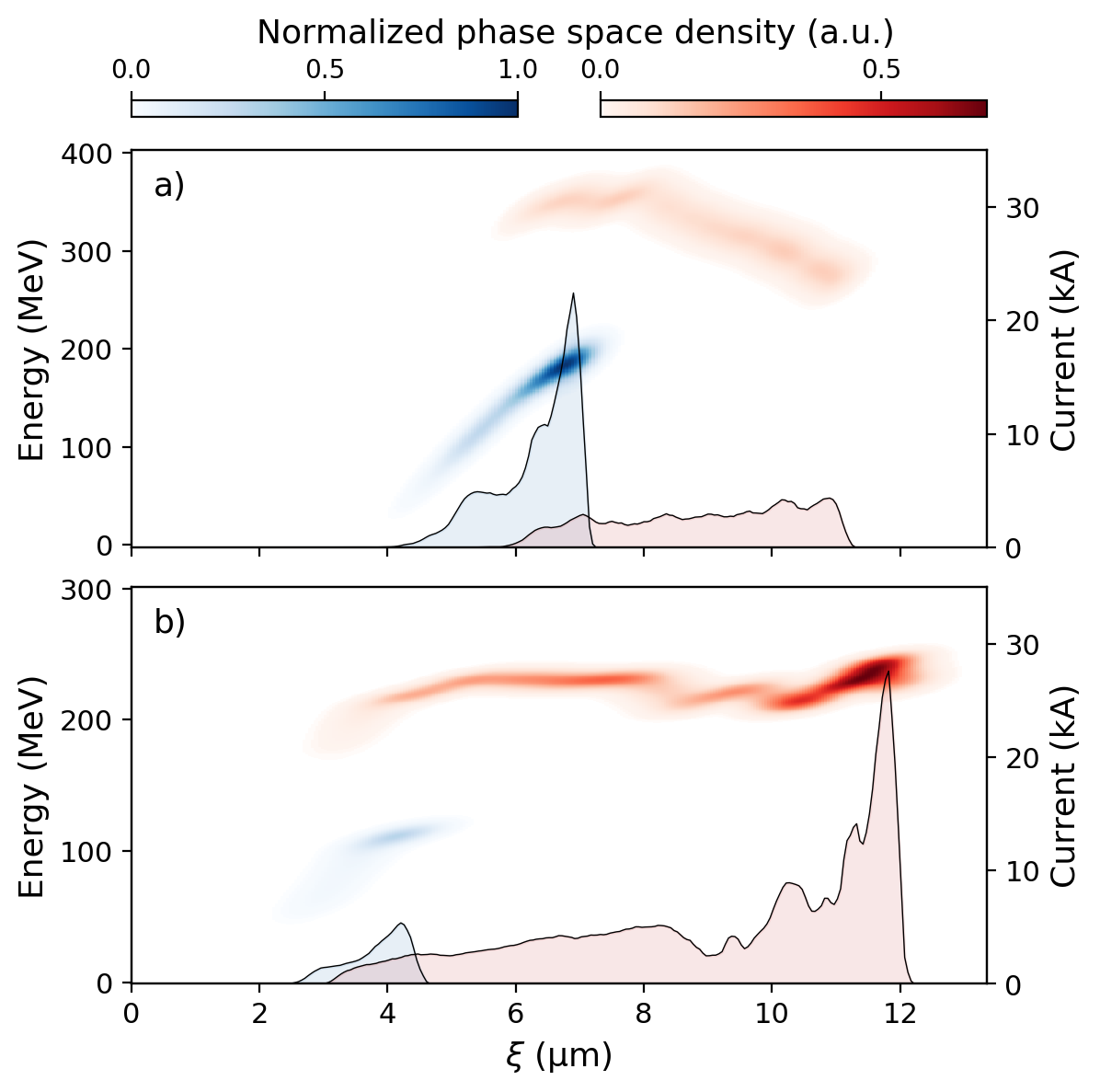}
\caption{PIC study of multiple torch injectors spaced by $\approx 2$ mm. Shown are longitudinal phase spaces of resulting witness ensembles for two cases: a), the first torch is comprised of $r_\mathrm{flat, \, 1} = 40\, \mathrm{\mu m}$, $l_\mathrm{ramp, \, 1} = 300\, \mathrm{\mu m}$ followed by $r_\mathrm{flat, \, 2} = 40\, \mathrm{\mu m}$, $l_\mathrm{ramp, \, 2} = 10\, \mathrm{\mu m}$ with reduced peak density $n_\mathrm{T, \, 2} = 0.3 \times n_{\mathrm{ch}}$. b), results for first torch with $r_\mathrm{flat, \, 1} = 40\, \mathrm{\mu m}$, $l_\mathrm{ramp, \, 1} = 80\, \mathrm{\mu m}$ followed by the same configuration as in a). Red and blue coloring reflect the origin of charge from the first and second injector stage, respectively.}
\label{fig:doubleTorch}
\end{figure}

The plasma torch scheme allows for generation of witness beams of various energies. This can be accommodated by varying the plasma accelerator length as in conventional plasma acceleration, but also by moving the position of the plasma torch within the plasma accelerator stage seamlessly. Furthermore, it is possible to employ multiple plasma torches to trigger multiple injection processes consecutively.  To demonstrate this, we conduct simulations including a second density spike downstream of the first plasma torch.
\Cref{fig:doubleTorch} presents two PIC simulations employing this double-torch configuration, each  with flat-top radius $r_\mathrm{flat} = 40\, \mathrm{\mu m}$ as shown in \cref{fig:rampScan} a) for a single torch. In both simulations, these plasma torch injectors are separated by $\Delta x \, \approx$ $2$ mm, leading to an energetic  spacing of roughly 100 MeV between the two witness beams.
The longitudinal phase space and associated slice currents are shown for the two double-torch configurations. In \cref{fig:doubleTorch} a), the first torch is using $n_\mathrm{T, \, 1} = 0.5 \times n_{\mathrm{ch}}$ and a long ramp $l_\mathrm{ramp, \, 1} = 300\ \mathrm{\mu m}$, which injects $46.3$ pC witness charge (red). The second torch has $n_\mathrm{T, \, 2} = 0.3 \times n_{\mathrm{ch}}$ and a short ramp $l_\mathrm{ramp, \, 2} = 10\ \mathrm{\mu m}$, and injects the second witness beam with $64.6$ pC (blue). 
The lower peak density of the second torch $n_\mathrm{T, \, 2} < n_\mathrm{T, \, 1}$, but its shorter ramp $l_\mathrm{ramp, \, 2} < l_\mathrm{ramp, \, 1}$, in this example produces a beam with higher charge, but shorter length and thus higher current than the beam from the long ramp injector.

The second case shown in \cref{fig:doubleTorch} b) varies the ramp length of the first torch to $l_\mathrm{ramp, \, 1} = 80\ \mathrm{\mu m}$, leaving other parameters unchanged.
In this configuration, the trapped beams contain charge of $\approx 160.1$ (first beam, red) and 16.4 pC (second beam, blue), respectively. Here, the  first beam substantially loads the wake and limits the amount of injected charge  of the second one. While in these cases the second beam is trapped at a later position within the wake (e.g. due to beam loading of the first beam), the relative position between the two can in principle be altered by controlling the plasma density level differences between the individual torch filaments. 
Further, the energy gap can be varied by tuning the position of both torches. This concept can be extended to even more torch injectors, each consuming only mJ-scale laser pulse energies, which is a minor fraction for the total energy budget of state-of-the-art laser systems. 
\newline
%that leads to formation of the production of such beam pairs has far reaching prospects in radio therapy, or for multi-color X-ray generation, which are of special interest, e.g., for medical imaging. 
%references 41-44 from Multichromatic Narrow-Energy-Spread Electron Bunches from Laser-Wakefield Acceleration with Dual-Color Lasers:
%[A. Marinelli et al., Phys. Rev. Lett. 111, 134801 (2013). [42] C. Paulus et al., J. Instrument. 8, P04003 (2013). [43] B. Dierickx et al., in European Optical Society symposium, Munchen (2009) pp. 16–18. [44] V. Malka et al., Nature Physics 4, 447 (2008)]

\bigskip
\section{Appendix D: Additional trapping doughnuts}

%\begin{center}
\begin{figure*}
 \includegraphics[width=1\textwidth]{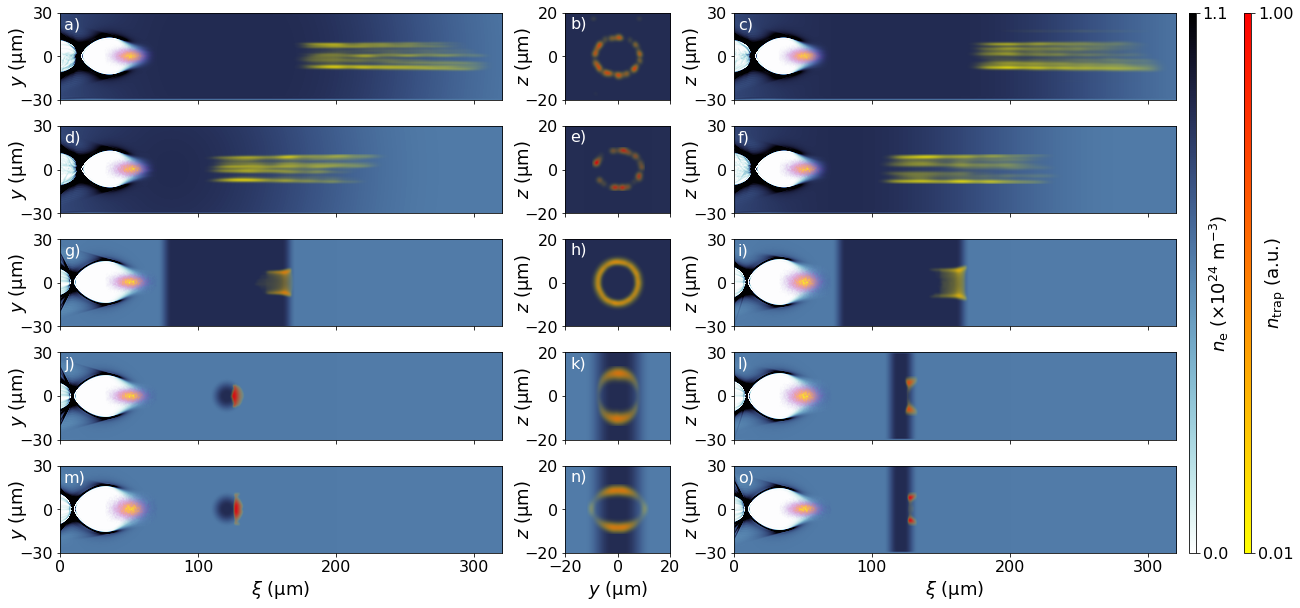}
\caption{Additional trapping doughnuts with similar PIC setting as in \cref{fig:donutsSymm,fig:donutsAsymm}. a)-c): $40 \, \mathrm{\mu}$m-wide flat-top radius and  $200 \, \mathrm{\mu}$m-long ramp, reduced longitudinal grid size to $0.1\ \mathrm{\mu}$m. 
d)-f): $2.5$ $\mathrm{\mu}$m-wide flat-top radius and  $200$ $\mathrm{\mu}$m-long ramp.  g)-i):  $40$ $\mathrm{\mu}$m-wide flat-top radius and  $10$ $\mathrm{\mu}$m-long ramp, asymmetric drive beam wider in $z$.  j)-l): $2.5$ $\mathrm{\mu}$m-wide flat-top radius and  $10$ $\mathrm{\mu}$m-long ramp, asymmetric drive beam wider in $z$.  m)-o): $2.5$ $\mathrm{\mu}$m-wide flat-top radius and  $10$ $\mathrm{\mu}$m-long ramp, asymmetric drive beam wider in $y$.}
\label{fig:addDonuts}
\end{figure*}
%\end{center}

\Cref{fig:addDonuts} presents further PIC simulations with similar setup as in \cref{fig:donutsSymm,fig:donutsAsymm}.
The first simulation repeats the one shown in \cref{fig:donutsSymm} d)-f) ($40$ $\mathrm{\mu}$m-wide flat-top radius and  $200$ $\mathrm{\mu}$m-long ramps) with reduced longitudinal grid size to $0.1\ \mathrm{\mathrm{\mu}}$m. The trapping volume displays a similar, irregular pattern.

In the second row, the torch consists of a $2.5$ $\mathrm{\mu}$m-wide flat-top radius and  $200$ $\mathrm{\mu}$m-long ramps in radial direction. Since $l_{\mathrm{ramp}} + r_{\mathrm{flat}} \gg r_{\mathrm{b}}$, the trapping volume resembles the closely related case shown in \cref{fig:donutsSymm} d)-f), which employs a $40$ $\mathrm{\mathrm{\mu}}$m-wide flat-top radius. Also, the trapped charge is  comparable.

The following simulations highlight the effect of an asymmetric drive beam ($\sigma_z = 1.5\times \sigma_y$) resulting in asymmetric blowout formation. In \cref{fig:addDonuts} g)-i), the torch consists of a $40$ $\mathrm{\mu}$m-wide flat-top radius and $10$ $\mathrm{\mu}$m-long ramps. Due to the asymmetric drive beam, the trapping doughnut in the $z y$-plane gets squeezed along the wider plane of the drive beam (blowout), and it gets compressed in the perpendicular direction, both by a factor of ${\sim}1.2$ relative to the symmetric case shown in \cref{fig:donutsSymm} a)-c). As a side note, the compressed shape re-appears for \cref{fig:donutsAsymm} g)-i), which employs an identical drive beam together with a torch density modulation in the $z y$-plane. 

\Cref{fig:addDonuts} j)-l) and \cref{fig:addDonuts} m)-o) present the effect of electron drive beams asymmetric in $z$ ($\sigma_z = 1.5\times \sigma_y$) and in $y$-direction ($\sigma_y = 1.5\times \sigma_z$), respectively, for a thin torch $l_{\mathrm{ramp}} + r_{\mathrm{flat}} \ll r_{\mathrm{b}}$. Here, the trapping doughnut gets cropped like shown in \cref{fig:donutsAsymm} a)-c) due to the narrow torch and, additionally, stretched (squeezed) along the wider (narrower) extent of the asymmetric drive beam. Furthermore, these two simulations emphasize that multiple effects can simultaneously be applied to the trapping volume.

\bibliographystyle{apsrev}
\bibliography{20200717_citations.bib}

\end{document}